\documentclass[a4paper,11pt]{article}
\pdfoutput=1 

\usepackage{jheppub} 

\usepackage{amsmath}
\usepackage{amssymb}
\usepackage{young}
\usepackage[vcentermath]{youngtab}
\usepackage{dsfont}
\usepackage{braket}
\usepackage{comment}
\usepackage{textgreek} 
\usepackage{simplewick}

\allowdisplaybreaks
\usepackage{mathrsfs}
\usepackage[mathscr]{euscript}

\numberwithin{equation}{section}

\def\be{\begin{equation}}
\def\ee{\end{equation}}
\newcommand{\LL}{\mathcal{L}^{\text
{L}}}
\newcommand{\LLm}{\mathcal{L}^{\text
{m}}}

\title{Three-point functions in AdS$_\mathbf{3}$/CFT$_\mathbf{2}$ holography}

\author{Andrea Dei,}
\author{Lorenz Eberhardt}
\author{and Matthias R.~Gaberdiel}

\affiliation{Institut f\"ur Theoretische Physik, ETH Z\"urich\\ Wolfgang-Pauli-Stra{\ss}e 27, 8093 Z\"urich, Switzerland}

\emailAdd{adei@itp.phys.ethz.ch}
\emailAdd{eberhardtl@itp.phys.ethz.ch}
\emailAdd{gaberdiel@itp.phys.ethz.ch}

\abstract{Recently, string theory on $\text{AdS}_3 \times \text{S}^3 \times \mathbb{T}^4$ with one unit of NS-NS flux ($k=1$) was argued to be exactly dual to the symmetric orbifold of $ \mathbb{T}^4$, and in particular, the full (unprotected) spectrum was matched between the two descriptions.  This duality was later extended to the case with higher NS-NS background flux for which the long string sector was argued to be described by the symmetric product orbifold of $(\mathcal{N}=4$ $\text{Liouville}) \times \mathbb{T}^4$. In this paper we study correlation functions for the bosonic analogue of this duality, relating bosonic string theory on $\text{AdS}_3 \times X$ to the symmetric orbifold of $\text{Liouville} \times X$. More specifically, we show that the low-lying null vectors of Liouville theory correspond to BRST exact states from the worldsheet perspective, and we demonstrate that they give rise to the expected BPZ differential equations for the dual CFT correlators. Since the structure constants of Liouville theory are uniquely fixed by these constraints, this shows that the seed theory of the dual CFT contains indeed the Liouville factor.}

\begin{document}
\maketitle
\flushbottom

\section{Introduction}

The AdS/CFT correspondence \cite{Maldacena:1997re} has been tremendously successful in giving deep insights into strongly coupled field theories on the one hand and various aspects of quantum gravity on the other. Many of its striking predictions have been verified independently over the years. However, despite this large amount of evidence, there is currently no proof of the correspondence. In fact, since the duality relates generically a weakly coupled description to one that is strongly coupled (and is therefore not accessible in terms of perturbative methods), this seems hard to circumvent. 

As always in physics, our understanding would benefit greatly from a \emph{solvable} example of the correspondence, and various attempts have been made over the years to find simplified systems where this may be achieved, see e.g.\ \cite{Gopakumar:1998ki,Klebanov:2002ja}. In order to find a fully-fledged solvable stringy duality, the case of the $\mathrm{AdS}_3/\mathrm{CFT}_2$ correspondence is particularily promising since both, $\mathrm{AdS}_3$ backgrounds (with pure NS-NS flux), and many $2d$ $\mathrm{CFT}$s, admit exact descriptions \cite{Maldacena:2000hw,DiFrancesco:1997nk}.

String compactifications on $\mathrm{AdS}_3$ have typically a high-dimensional moduli space,  and the same is true for the dual CFTs. For example, it has long been believed that string theory on any $\mathrm{AdS}_3 \times \mathrm{S}^3 \times \mathbb{T}^4$ background  is dual to a $2d$ CFT that lies on the same moduli space as the symmetric product orbifold $\mathrm{Sym}^N(\mathbb{T}^4)$, see e.g.\ \cite{David:2002wn} for a review. However, until recently it was not clear which precise string background corresponds to the symmetric product orbifold itself. As a consequence one could only compare protected quantities (that are independent of the specific point in moduli space), such as the structure of the complete moduli space \cite{Dijkgraaf:1998gf, Larsen:1999uk}, the BPS spectrum and the associated supersymmetric indices \cite{Seiberg:1999xz, Maldacena:1999bp, Argurio:2000tb, deBoer:1998us}, as well as extremal correlators \cite{Gaberdiel:2007vu, Dabholkar:2007ey}. 
\smallskip

It was recently proposed that string theory on $\mathrm{AdS}_3 \times \mathrm{S}^3 \times \mathbb{T}^4$ with exactly one unit ($k=1$) of NS-NS flux (and no R-R flux) is exactly dual to the symmetric product orbifold of $\mathbb{T}^4$ in the large $N$ limit \cite{ Gaberdiel:2018rqv, Eberhardt:2018ouy}. This theory is believed to be `tensionless' \cite{Gaberdiel:2014cha,Gaberdiel:2017oqg,Ferreira:2017pgt}, which is consistent with why one may hope to have weakly coupled (perturbative) descriptions on both sides of the correspondence. The main support for the proposal comes from the fact that the complete (unproteced) spectrum matches between the two descriptions \cite{ Gaberdiel:2018rqv, Eberhardt:2018ouy}, see also \cite{Giribet:2018ada} for a related analysis. In addition, it was shown in \cite{Eberhardt:2019qcl}, generalising slightly techniques of \cite{Giveon:1998ns,deBoer:1998gyt,Kutasov:1999xu}, that the symmetry algebra (the spectrum generating algebra) also agrees. It is usually believed that a CFT is uniquely characterised by its spectrum, together with the complete set of structure constants, i.e.\ the set of all $3$-point functions. Given that the spectrum agrees, it therefore only remains to check that the structure constants of the symmetric orbifold are correctly reproduced from the stringy worldsheet theory. 
\medskip

In this paper, we shall take a first step towards proving this claim. Instead of dealing with the above setup, we shall consider the generalisation to generic pure NS-NS flux, where the long string sector is conjectured to be dual to \cite{Eberhardt:2019qcl}\footnote{ We should mention that the definition of the symmetric orbifold for Liouville theory is a bit subtle since one has to add the vacuum to Liouville theory, see \cite{Eberhardt:2019qcl} for more details. This prescription is dictated by the requirement that the worldsheet spectrum is correctly reproduced.}
\be 
\text{Sym}^N\big(\big(\text{$\mathcal{N}=4$ Liouville theory with $c=6(k-1)$}\big) \times \mathbb{T}^4\big)\ . \label{eq:introN4 Liouville sym}
\ee
Here, $k$ denotes the amount of NS-NS flux of the $\mathrm{AdS}_3 \times \mathrm{S}^3 \times \mathbb{T}^4$ background. Note that in the minimal flux case (i.e.\ for $k=1$) the Liouville factor becomes trivial, and the theory reduces to the symmetric orbifold of $\mathbb{T}^4$. 

Actually, in order to avoid some of the technical complications coming from supersymmetry, we shall mainly consider the bosonic analogue, which was also studied in \cite{Eberhardt:2019qcl}: it postulates that the long string sector of string theory on $\mathrm{AdS}_3 \times X$  is equivalent to 
\be 
\text{Sym}^N\left(\left(\text{Liouville with $c^{\text L}=1+\frac{6(k-3)^2}{k-2}$} \right)\times X \right)\ ,\label{eq:dual CFT}
\ee
see also \cite{Argurio:2000tb}. We expect our analysis to work similarly in the supersymmetric case, but the details seem to be complicated, see also the discussion in Section~\ref{sec:conclusions}.

The main advantage of dealing with bosonic Liouville theory is that its correlation functions are \emph{known} to be determined by the constraints that arise from the decoupling of null vectors \`a la \cite{Belavin:1984vu}. Indeed, following Teschner \cite{Teschner:1995yf}, one can use the BPZ equation associated to a certain null vector together with crossing symmetry to deduce the famous DOZZ formula \cite{Dorn:1992at,Dorn:1994xn,Zamolodchikov:1995aa} for the structure constants of Liouville theory. Thus the `seed' theory of the dual CFT is indeed Liouville theory (of the appropriate central charge) provided that we can show that the relevant null vector is also null from a worldsheet perspective. It is the aim of this paper to establish exactly this: more precisely, we shall show that 
\begin{verse} 
\hspace{0.5cm} \textit{the level one and level two null vectors of Liouville theory correspond to BRST exact states in the worldsheet theory.}
\end{verse}
Since BRST exact states decouple in physical correlation functions, this leads to differential equations for the worldsheet correlators. We have also checked (see Section~\ref{sec:BPZ-from-the-worldsheet}) that they lead indeed to the correct BPZ equations of the spacetime CFT using this  worldsheet viewpoint. 
\smallskip

While this effectively shows that all untwisted correlation functions of the symmetric orbifold --- these are the ones that are determined by those of the seed theory --- match, unfortunately our methods do not suffice yet to prove an analogous statement for all correlators (including twisted sector states). In particular, while we can show that also the twisted generalisations of these null vectors vanish from a worldsheet perspective (this will be done in Section~\ref{sec:twisted}), it is not clear to us whether this will suffice\footnote{We suspect that it does not. In any case, little seems to be known about how to constrain the symmetric orbifold correlators using null vector techniques.} to deduce all structure constants using some generalisation of the Teschner argument. 
\medskip

Besides providing another strong piece of evidence for the correspondence, our result also has some other consequences. It gives rise to some exact correlators of the $\mathrm{SL}(2,\mathds{R})_k$ WZW model in spectrally flowed sectors; to our knowledge, these correlators were previously out of reach, see however \cite{Teschner:1997ft, Teschner:1999ug, Giribet:2001ft, Ribault:2005ms, Giribet:2007wp, Maldacena:2001km} for partial results. Interestingly, the analytic structure of the spectrally flowed correlators turns out to be quite different from their unflowed analogues: they are essentially given by the DOZZ formula of Liouville theory \cite{Dorn:1992at, Dorn:1994xn, Zamolodchikov:1995aa}, with the parameter $b$ given by eq.~\eqref{eq:b string}.
\bigskip

The paper is organised as follows. In Section~\ref{sec:bosonic strings} we fix our conventions and review some important concepts about strings on $\mathrm{AdS}_3$. We also explain in detail how the $x$-basis, i.e.\ the position basis of the dual CFT, appears from a worldsheet perspective. Section~\ref{sec:CFTnull=BRSTexact} is the core of the paper where we show our main result, namely that the level one and level two null vectors of Liouville theory are BRST exact on the worldsheet. This is generalised to the low-lying null vectors in twisted sectors in Section~\ref{sec:twisted}. In Section~\ref{sec:BPZ-from-the-worldsheet} we show that the corresponding decoupling equations lead indeed to the expected BPZ equation for the dual CFT, and we conclude in Section~\ref{sec:conclusions} with some outlook. There are three appendices: some background material about $\mathrm{AdS}_3$ is collected in Appendix~\ref{app:conventions}, while Appendices~\ref{app:LHS-RHS-coeff} and \ref{app:generic-w} contain some of the technical details of the computations in Sections~\ref{sec:CFTnull=BRSTexact} and \ref{sec:twisted}, respectively.

\section{Bosonic strings on \texorpdfstring{AdS$_\mathbf{3}$}{AdS3}} \label{sec:bosonic strings}

In this paper we shall be analysing bosonic string theory on ${\rm AdS}_3$. From a worldsheet perspective, this background will be described in terms of an  $\mathrm{SL}(2,\mathds{R})_k$ WZW model, following \cite{Maldacena:2000hw}. We shall work essentially with the conventions of \cite{Eberhardt:2019qcl} which we review for the benefit of the reader in Appendix~\ref{app:conventions}. There are two important issues that will play an important role in our analysis, and we shall explain them in some detail here.

\subsection{The DDF algebra}

The physical states of string theory on $\mathrm{AdS}_3 \times X$ correspond to Virasoro primary states on the worldsheet. It is convenient to describe them in terms of the so-called DDF operators \cite{DelGiudice:1971yjh} that map physical states to physical states. For the case of ${\rm AdS}_3$, the DDF operators were first constructed in  \cite{Giveon:1998ns} and it was shown \cite{Giveon:1998ns, Eberhardt:2019qcl} that they act naturally on the continuous (i.e.~long string) part of the Hilbert space.\footnote{Some of the DDF operators also act on the discrete (i.e.~short string) part of the Hilbert space, namely the integer-moded operators, see also footnote \ref{footnote:fractional moded}. Since the short string sector does not appear in \eqref{eq:dual CFT}, we will focus on the long string sector here.} They include in particular the Virasoro generators of the spacetime CFT\footnote{Here and in the following, we absorb a factor of $2\pi i$ in the definition of the contour integral.}
\be 
\mathcal{L}_m = \oint_0 \mathrm{d}z\ \left((1-m^2) \gamma^m J^3 + \dfrac{m(m-1)}{2} \gamma^{m+1} J^+ + \dfrac{m(m+1)}{2}\gamma^{m-1} J^- \right)(z)  \ .
\label{eq:spacetime-Vir}
\ee
Here $J^a$ are the $\mathfrak{sl}(2,\mathds{R})_k$ currents, while $\gamma$ is the field that appears in the Wakimoto representation of $\mathfrak{sl}(2,\mathds{R})_k$, see Appendix~\ref{app:conventions} for our conventions. We will always denote the spacetime generators (i.e.\ the DDF operators) by curly letters, while worldsheet generators are described by straight letters. We mention in passing that the global M\"obius generators simply reduce to 
\be \label{zeromodes}
\mathcal{L}_0 = \oint_0 \mathrm{d}z \ J^3(z) = J^3_0  \ , \quad\ \ \mathcal{L}_{-1} = \oint_0 \mathrm{d}z \ J^+(z) = J^+_0  \ , \quad\ \ \mathcal{L}_1 = \oint_0 \mathrm{d}z \ J^-(z) = J^-_0 \ . 
\ee
The DDF operators in eq.~\eqref{eq:spacetime-Vir} satisfy the spacetime Virasoro algebra
\be 
[\mathcal{L}_m, \mathcal{L}_n] = (m-n) \mathcal{L}_{m+n} + \tfrac{k}{2}\,  \mathcal{I}m(m^2-1) \delta_{m+n,0} \ , 
\ee
where the central element $\mathcal{I}$ takes the value $w$ in the sector with spectral flow $w$  \cite{Giveon:1998ns, Eberhardt:2019qcl}. The total central charge is therefore $c=6kw$, and therefore seems to depend on $w$. However, as explained in \cite{Eberhardt:2019qcl}, this fits perfectly with the interpretation that the dual CFT is a symmetric orbifold of a seed theory with $c_{\text{seed}}=6k$, provided that $w$ is the length of a single cycle twist.\footnote{As explained in \cite{Eberhardt:2019qcl}, the modes of $\mathcal{L}_n$ take values in $n \in \frac{1}{w} \mathds{Z}$ in the $w$-twisted sector. After redefining them in terms of the generators of the seed theory, they lead to a Virasoro algebra with central charge $c_{\text{seed}}=6k$ for all $w$, see the discussion in Section~2.5.1 of \cite{Eberhardt:2019qcl}. \label{footnote:fractional moded}}

In \cite{Eberhardt:2019qcl}  the full set of DDF operators was analysed systematically, and it was found that they generate the spacetime CFT symmetry algebra
\be \label{spacetimechiral}
\left(\text{Virasoro with $c^{\text L}=1+\frac{6(k-3)^2}{k-2}$}\right) \times \left(\text{chiral algebra of $X$}\right)\ .  
\ee
Together with the analysis of the spectrum from the long string sector this led to the conjecture  \cite{Eberhardt:2019qcl} that the corresponding dual CFT is the symmetric product orbifold of ${\rm Liouville} \times X$, see eq.~(\ref{eq:dual CFT}). In the following we shall refer to fields coming from the $X$ part as `matter fields'. 

We shall mainly be interested in the Virasoro generators of the Liouville part. They can be obtained from the total Virasoro generators in (\ref{eq:spacetime-Vir}) upon subtracting out the spacetime Virasoro generators $\LLm_n$ associated to the matter CFT $X$,
\be \label{LiouvilleL}
\LL_n = \mathcal{L}_n - \LLm_n \ ,  
\ee
where \cite{Eberhardt:2019qcl}
\be
\mathcal{L}_m^\text{m} = \oint_0 \mathrm{d}z\ \big((\partial \gamma)^{-1} \gamma^{m+1} T^\text{m}\big)(z)+\frac{c^\text{m}}{12} \oint_0 \mathrm{d}z\ \gamma^{m+1} \left(\frac{3}{2}(\partial^2\gamma)^2 (\partial \gamma)^{-3}-\partial^3 \gamma(\partial \gamma)^{-2}\right)\ .\label{eq:matter Virasoro spacetime}
\ee
These generators satisfy a Virasoro algebra with central charge 
\be \label{cm}
c^\text{m}=26-\frac{3k}{k-2} \ ,
\ee
and the central charge of the Liouville factor is therefore 
\be 
c^\text{L}=c_{\text{seed}}-c^{\text{m}}=1+\frac{6(k-3)^2}{k-2}\ , \label{eq:Liouville-central-charge}
\ee
see eq.~(\ref{spacetimechiral}).

\subsection[The vertex operators in the \texorpdfstring{$x$}{x}-basis]{The vertex operators in the $\boldsymbol{x}$-basis} \label{subsec:x basis}

The main aim of the paper is to show that the null vectors of this spacetime Liouville theory are also null from the worldsheet perspective, and to study the resulting constraints on correlation functions. 
On the worldsheet, it will hence be important to keep track of the positions $x_i$ where the fields are inserted in the spacetime CFT. In addition, the vertex operators will also depend on $z_i$, the coordinate on the worldsheet that is integrated in order to obtain the string theory correlators.

We are interested in describing the worldsheet vertex operator that corresponds to a spacetime quasiprimary state of conformal weight $h$, i.e.\ a state satisfying
\be
\mathcal{L}_0 \, |h\rangle  = J^3_0 \,  |h\rangle = h \,  |h\rangle  \ , \qquad \hbox{and} \qquad \mathcal{L}_1 \, |h\rangle  =  J^-_0 \,  |h\rangle =0 \ ,
\ee
where we have used the identifications (\ref{zeromodes}). Via the usual state field correspondence, the state should be identified with the field acting on the vacuum at zero; in the present context, there are two coordinates at play, and we should therefore set
\be\label{fieldstate}
\left. V_h(x;z) |0\rangle \right|_{x=0 , \,  z=0 } = |h\rangle \ . 
\ee
Furthermore, since $\mathcal{L}_{-1}$ and $L_{-1}$ are the translation operators in spacetime and on the worldsheet, respectively, the vertex operator at generic positions is defined via 
\be\label{doubletranslation}
\mathrm{e}^{\zeta {L}_{-1}} \,  \mathrm{e}^{y\mathcal{L}_{-1}} \, V_h(x;z) \, \mathrm{e}^{-y\mathcal{L}_{-1}}\,  \mathrm{e}^{-\zeta {L}_{-1}}\, =V_h(x+y;z+\zeta)\ .
\ee
It follows that the global $\mathfrak{sl}(2,\mathds{R})$ generators act on spacetime quasi-primary vertex operators as differential operators, i.e.\ we have 
\begin{subequations} \label{xbasis}
\begin{align}
[J^+_0, V_h(x;z)]&=\frac{\partial}{\partial x} V_h(x;z) \ , \\
[J^3_0, V_h(x;z)]&=\left(h + x \frac{\partial}{\partial x} \right)V_h(x;z) \ , \\
[J^-_0, V_h(x;z)]&= \left(2 h x +x^2 \frac{\partial}{\partial x} \right) V_h(x;z) \ . 
\end{align}
\end{subequations}
By integrating up these equations we then find 
\begin{align} 
U(g) V_{h}(x;z) U(g)^{-1}&=M_g'(x)^h \ V_h(M_g(x);z)\ , \\
M_g(x)&=\frac{a x+b}{c x+d}\ ,\qquad g=\begin{pmatrix}
a & b \\ c & d
\end{pmatrix} \in \mathrm{SL}(2,\mathds{R})\ ,
\end{align}
where $U(g)$ is an exponential in $J^a_0$ that describes the action of the group element on the field $V_h(x;z)$; this formula generalises \eqref{doubletranslation}. 

Since the zero modes $J^a_0$ act naturally on $x$, we can think of $x$ as labelling the different states in the $\mathfrak{sl}(2,\mathds{R})$ representation (of the affine highest weight states). This way of parametrising the vertex operators on the worldsheet is therefore often referred to as the `$x$-basis'. We should mention in passing that the corresponding state $|h\rangle$ in eq.~(\ref{fieldstate}) lives then in the so-called `$m$-basis' of  $\mathfrak{sl}(2,\mathds{R})$, in which the $J^3_0$ generator acts diagonally. Thus both the $m$-basis and the $x$-basis appear naturally in this context. 
\smallskip

While this discussion may seem somewhat unnecessarily pedantic, it is actually important for understanding the behaviour of these vertex operators under spectral flow. We will define spectral flow to act on the states at $(x;z)=(0;0)$ (that live naturally in the $m$-basis). This will then fix the definition of the corresponding vertex operators for arbitrary $(x;z)$, using (\ref{doubletranslation}). In particular, the `direction' of spectral flow will depend on $x$, since the $x$-translation operator $J^+_0$ does not commute with the spectral flow direction $J^3$, see also \cite{Eberhardt:2018ouy}. 

As a consequence there are two natural fusion rules one may define (and this has led to some confusion in the literature). The fusion rules that will be relevant for the description of the spacetime CFT refer to the conditions under which the $n$-point functions (and more specifically the $3$-point functions)
\be 
\langle V_{h_1}(x_1;z_1) \cdots V_{h_n}(x_n;z_n)\rangle 
\ee
do not vanish for generic $x_i$ and $z_i$. As was explained in \cite{Maldacena:2001km, Eberhardt:2018ouy, LorenzMatthiasRajesh}, they take the form 
\be 
[w_1]  \times_x [w_2]  \cong \bigoplus_{\genfrac{}{}{0pt}{}{w=|w_1-w_2|+1}{w + w_1 + w_2 \equiv 1 \bmod 2}}^{w_1+w_2-1} [w]\  ,
\ee
where $[w]$ stands for representations that arise in the $w$'th spectrally flowed sector. These fusion rules are sometimes referred to as the $x$-basis fusion rules. 

On the other hand, they are to be contrasted with the usual worldsheet CFT fusion rules where one would not introduce the $x$ dependence as above, and therefore insert all vertex operators at $x_i=0$;\footnote{From the viewpoint of \cite{LorenzMatthiasRajesh}, this corresponds to the singular configuration where the branch points all coincide. Then the covering map is in some sense trivial, and the order of its branch point is the sum of the orders of the individual branch points.} then the fusion rules obey instead \cite{Maldacena:2001km}
\be 
[w_1]  \times_m [w_2]  \cong \bigoplus_{w=w_1+w_2-1}^{w_1+w_2+1} [w]\  .
\ee
These fusion rules are sometimes referred to as the $m$-basis fusion rules.

\section{The Liouville null states are BRST exact on the worldsheet}
\label{sec:CFTnull=BRSTexact}

This section is the core of the paper: we want to show that the null vectors of the spacetime Liouville theory are BRST exact on the worldsheet and hence vanish inside physical correlation functions. Among other things, this will confirm that the spacetime dual contains indeed a Liouville theory of the correct central charge. The Liouville null fields arise in the untwisted sector of the symmetric orbifold; they will therefore correspond to vertex operators from the $w=1$ sector of the worldsheet. We shall comment on the case with $w>1$ further below, see Section~\ref{sec:twisted}.
\smallskip

For the following it will be convenient to add a total derivative to the integrand in \eqref{eq:spacetime-Vir}, and to rewrite the spacetime Virasoro modes as 
\be 
\mathcal{L}_m=\oint_0 \mathrm{d}z\ \big((m+1) \gamma^m J^3-m \gamma^{m+1} J^+\big)(z) \ .
\label{eq:simplified-spacetime-Vir}
\ee
This will make some of the subsequent calculations easier. 

\subsection{Null states in Liouville theory}
\label{sec:null-states-CFT}

Let us start by recalling the structure of null vectors in Liouville theory. Consider a Virasoro primary state $\ket{h}$ of conformal weight $h$ with respect to $\LL_0$. At level one, the only null vector is the descendant of the vacuum,
\be\label{N1}
\ket{N_1} = \LL_{-1}\ket{0} \ . 
\ee
Indeed the latter is annihilated by the action of all positive Virasoro modes. At level two we start with the ansatz 
\be \label{N2}
\ket{N_2} = \left(\LL_{-2} + a (\LL_{-1})^2\right)\ket{h_*}
\ee
and solve for $a$ and $h_*$ by imposing
\be 
\LL_1 \ket{N_2} = 0 \ , \qquad \text{and} \qquad \LL_2 \ket{N_2} = 0 \ . 
\ee
One finds
\begin{subequations}
\begin{align} 
\mathcal{L}_1 \ket{N_2} & = \bigl(3 + (4 h_* + 2)a \bigr) \mathcal{L}_{-1} \ket{h_*} \ , \\
\mathcal{L}_2 \ket{N_2} & = \left((4 + 6a) h_* + \frac{c}{2} \right) \ket{h_*} \ , 
\end{align}
\end{subequations}
and hence
\begin{subequations} \label{N2cond}
\begin{align}
a & = \dfrac{c - 13 \pm \sqrt{(c-25)(c-1)}}{12} \ ,  \label{eq:a level 2 deg}\\
h_* & = \dfrac{5 - c \pm \sqrt{(c-25)(c-1)}}{16} \ . \label{eq:hstar level 2 deg}
\end{align}
\end{subequations}
We will reserve the symbol $h_*$ in the following for the conformal weight of this degenerate field.
By rewriting $c$ as $c = 1 + 6 Q^2$ with $Q = b + b^{-1}$, this becomes 
\be \label{3.7}
a = b^{\pm 2}\ , \qquad h_* = - \left( \dfrac{1}{2} + \dfrac{3}{4} b^{\mp 2} \right) \ . 
\ee
For the theory at hand we have furthermore $c=c^\text{L}$ (see eq.~\eqref{eq:Liouville-central-charge}), and thus
\be 
 b^2=\left(\frac{k-3 \pm \sqrt{k^2-10k+17}}{2\sqrt{k-2}}\right)^2\ .\label{eq:b string}
\ee

\subsection{BRST exact states in string theory}
\label{sec:pure-gauge}

The main aim of this section is to show that the above null states are BRST exact from the viewpoint of the worldsheet. In bosonic string theory, BRST exact states are states that are both physical and orthogonal to any physical state. They can be shown to be always of the form, see e.g.\ \cite{Green:1987sp}
\be \label{BRSTex}
\ket{\psi} = L_{-1} \ket{\chi_1} + \left( L_{-2} + \dfrac{3}{2} L_{-1}^2 \right) \ket{\chi_2} \ ,
\ee
where 
\begin{subequations} 
\begin{align}
& L_m \ket{\chi_1} = 0 \qquad  m>0 \ , \qquad  && L_0 \ket{\chi_1} = 0 \ , \label{eq:pure-gauge-condition a}\\
& L_m \ket{\chi_2} = 0 \qquad  m>0 \ , \qquad  && L_0 \ket{\chi_2} = -\ket{\chi_2} \ . \label{eq:pure-gauge-condition b}
\end{align}
\end{subequations}

\subsection{CFT null states as BRST exact states}

With these preparations at hand, our task is now clear: we want to show that the null states $|N_1\rangle$ and $|N_2\rangle$ of Liouville theory, see eqs.~(\ref{N1}) and (\ref{N2}) with (\ref{N2cond}), respectively, are described by BRST exact operators from the worldsheet perspective, i.e.\ correspond to worldsheet states of the form (\ref{BRSTex}).

\subsubsection[The level one null vector \texorpdfstring{$|N_1\rangle$}{|N1>}]{The level one null vector $\boldsymbol{|N_1\rangle}$}

Let us start with the relatively simple case of $|N_1\rangle$. At level one, we will only need the first term in (\ref{BRSTex}), i.e.\ we claim that there exists a state $\ket{\chi_1}$ with $L_n \ket{\chi_1} = 0$ for $n\geq 0$ such that 
\be 
\LL_{-1} \ket{0} = L_{-1} \ket{\chi_1} \ . 
\label{eq:thesis-lev-one}
\ee 
To start with we need to identify the state $\ket{0}$ that describes the spacetime vacuum from the worldsheet perspective. This should arise from the $w=1$ sector, and we claim that the relevant state is the spectrally flowed image of the affine highest weight $|j,m\rangle^{(1)}$ --- the superindex indicates in which spectrally flowed sector this state is to be taken in --- with 
\be
m + \frac{k}{2} = 0  \quad \Longrightarrow \quad m = - \frac{k}{2} \ , 
\ee
so that $J^3_0 = \tilde{J}^3_0 + \frac{k}{2}$ has eigenvalue $0$. (Here and in the following we shall always use the convention that the modes before spectral flow are labelled by a tilde, and that $\tilde{J}^3_0\,  |j,m\rangle^{(1)} = m \, |j,m\rangle^{(1)}$, see Appendix~\ref{app:conventions} for more details.) Furthermore, the mass-shell condition requires that 
\be
1 = - \frac{j(j-1)}{k-2} - m - \frac{k}{4} = - \frac{j(j-1)}{k-2} + \frac{k}{4} \ , 
\ee
and thereby fixes $j$ to be 
\be\label{3.14}
j = \frac{1}{2} \pm \frac{k-3}{2} \ , \qquad \text{i.e.} \qquad j = \frac{k}{2}-1  \quad \hbox{or} \quad j = 2-\frac{k}{2} \ . 
\ee
We should mention in passing that this state is strictly speaking not part of the WZW spectrum (since $j \not\in \tfrac{1}{2} + i \, \mathds{R}$ but the representation we are looking at is a continuous representation since $m-j<0$ for $k>1$). This is immaterial for the following --- it just reflects that also the vacuum is not part of the Liouville spectrum. In fact, none of the null vectors of Liouville that allow one to deduce the DOZZ formula \`a la \cite{Teschner:1995yf}, appear in the actual Liouville spectrum. 
\smallskip

The left-hand-side of (\ref{eq:thesis-lev-one}) is now
\be
\LL_{-1} \ket{0} = \left( \mathcal{L}_{-1} - \LLm_{-1} \right) \ket{0} = \mathcal{L}_{-1} \ket{0} =   J_0^+ \ket{j,m}^{(1)} =  \tilde J_{-1}^+ \ket{j,m}^{(1)}  \ , 
\ee
where we have first used that the vacuum is also annihilated by the matter part of the Virasoro algebra. Here the parameters take the values $m=-\frac{k}{2}$,  while $j$ is given by eq.~(\ref{3.14}). Finally, in the last step we have utilised that this state arises from the $w=1$ sector.  It is easy to see that this state is then also physical, as it has to be. 

It remains to show that this state is of the form of the right-hand-side of (\ref{eq:thesis-lev-one}). We make the ansatz that 
$\ket{\chi_1} \propto |j,m+1\rangle^{(1)}$, again with $w=1$. (This state has then $L_0$ eigenvalue zero and is annihilated by the positive $L_n$ modes.) Applying now the Sugawara construction (in the $w=1$ sector) to this state, we find 
\begin{align}
L_{-1} \ket{j,m+1}^{(1)} &= \left(\tilde{L}_{-1} - \tilde{J}^3_{-1} \right) \ket{j,m+1}^{(1)} \\
&= \left( \dfrac{1}{k-2} \big(-2 \tilde J^3_{-1} \tilde J^3_0 + \tilde J_{-1}^+ \tilde J_0^- + \tilde J_{-1}^- \tilde J_0^+ \big) - \tilde J^3_{-1} \right) \ket{j,m+1}^{(1)} \\ 
&=  - \dfrac{2m+k}{k-2}\, \tilde{J}_{-1}^3 \ket{j,m+1}^{(1)} + \dfrac{m + 1 -j}{k-2} \, \tilde J_{-1}^+ \ket{j, m}^{(1)} \nonumber\\
&\qquad  + \dfrac{m+j+1}{k-2}\,  \tilde J_{-1}^- \ket{j, m+2}^{(1)}\ , 
\end{align}
where we have used the representation of zero modes given in \eqref{eq:tilded-zero-modes}. For $m = -\frac{k}{2}$ and $j = \frac{k}{2} - 1$ this then simplifies to 
\be 
L_{-1} \ket{\tfrac{k}{2} - 1,-\tfrac{k}{2}+1}^{(1)} = - \tilde J_{-1}^+\ket{\tfrac{k}{2} - 1, -\tfrac{k}{2}}^{(1)} =  - J_0^+\ket{\tfrac{k}{2} - 1, -\tfrac{k}{2}}^{(1)} \ , 
\ee
hence proving \eqref{eq:thesis-lev-one} with $\ket{\chi_1} = - \ket{\tfrac{k}{2}-1, -\tfrac{k}{2}+1}^{(1)}$. Thus we have shown that the $\mathcal{L}_{-1} |0\rangle$ state of the dual CFT is indeed BRST exact on the worldsheet, and hence vanishes in physical correlation functions. 

\subsubsection[The level two null vector \texorpdfstring{$|N_2\rangle$}{|N2>}]{The level two null vector $\boldsymbol{|N_2\rangle}$}
\label{sec:level-two-BRST-exact}

Now we want to repeat this analysis for the level two null vector, i.e.\ we want to show that 
\be 
\ket{N_2} = \Bigl( \LL_{-2} + b^{\pm 2} (\LL_{-1})^2 \Bigr) \ket{h_*} = \Bigl(L_{-2} + \dfrac{3}{2} L_{-1}^2 \Bigr) \ket{\chi_2} + L_{-1} \ket{\chi_1} \
\label{eq:thesis-lev-two}
\ee
for some suitable states $\ket{\chi_1}$, $\ket{\chi_2}$. As we will see, this will also fix $b$ as a function of $k$, and thereby confirm (\ref{eq:b string}).
We again identify the primary state $\ket{h_*}$ with an affine primary at $w=1$ on the worldsheet, 
\be 
\ket{h_*} = \ket{j,m}^{(1)} \ , \qquad  h_* = m + \frac{k}{2} \ , \qquad -\dfrac{j(j-1)}{k-2} - m - \frac{k}{4} = 1 \ ,
\label{eq:mass-shell-lev-two}
\ee
where the second relation just comes from the fact that $\mathcal{L}_0 = J^3_0$ (together with the fact that we assume that the state is the ground state with respect to the `matter' part), while the last condition is the mass-shell condition. 
By charge considerations, the only candidate for $\ket{\chi_2}$ is $\ket{\chi_2}\propto |j,m+2\rangle^{(1)}$, while for $\ket{\chi_1}$ we make the ansatz 
\begin{multline} \label{chi1}
\ket{\chi_1} = 4 (j - m - 3) x_1 \tilde{J}_{-1}^3 \ket{j, m+2}^{(1)} + x_2 \tilde{J}_{-1}^+ \ket{j, m+1}^{(1)} \\
+ (4 + k + 2m) x_1 \tilde{J}_{-1}^- \ket{j, m+3}^{(1)} \ ,
\end{multline}
where at this stage $x_1, x_2$ are arbitrary parameters (that may depend on $j$, $k$ and $m$). Here the coefficients in (\ref{chi1}) have been chosen so that $\ket{\chi_1}$ is Virasoro primary on the worldsheet. 

\paragraph{The DDF descendant}

In order to proceed we first need to calculate the left-hand-side of (\ref{eq:thesis-lev-two}). As explained above, see eq.~(\ref{LiouvilleL}), the Liouville Virasoro operators are given by 
\be 
\LL_n = \mathcal{L}_n - \LLm_n \ , 
\label{eq:LL=L-LLm0}
\ee
where $\mathcal{L}_m$ and $\LLm_m$ are defined in \eqref{eq:spacetime-Vir} and \eqref{eq:matter Virasoro spacetime}, respectively. 
Since we consider the matter CFT to sit in the vacuum, for the $-1$ mode this does not make a big difference
\be
\LL_{-1} \ket{h_*} = \mathcal{L}_{-1} \ket{h_*} = J_0^+ \ket{h_*} = J_0^+ \ket{j,m}^{(1)}   \ . 
\ee 
However, the matter CFT does contribute to the $-2$ mode since 
\begin{align}
\LLm_{-2} \ket{j,m}^{(1)}  
& = \oint_0 \mathrm{d}z \left( \dfrac{T^{\text{m}}}{z \tilde \gamma (\tilde \gamma + z \partial \tilde \gamma)} \right)(z) \ket{j,m}^{(1)}\nonumber\\ 
& \qquad + \dfrac{c^{\text{m}}}{12} \oint_0 \frac{\mathrm{d}z}{z}  \ \tilde \gamma^{-1}  \left( \dfrac{3(2\partial\tilde{\gamma}+z\partial^2 \tilde{\gamma})^2}{2(\tilde{\gamma}+z\partial \tilde{\gamma})^{3}}   - \frac{3\partial^2\tilde{\gamma}+z\partial^3 \tilde{\gamma}}{ (\tilde{\gamma}+z\partial \tilde{\gamma})^{2}} \right)(z) \ket{j,m}^{(1)} \ .  \label{eq:matter Virasoro level 2}
\end{align}
Here, we have written the result in the sector before spectral flow, by rewriting $\gamma(z)$ in terms of $\tilde{\gamma}(z)=z^{-1}\gamma(z)$, which is regular at the origin.
Since both $\tilde{\gamma}$ and $T^{\text{m}}$  only produce regular terms when acting on $\ket{j,m}^{(1)}$, we find 
\begin{align}
\LLm_{-2} \ket{j,m}^{(1)} & =  \left( \tilde \gamma^{-2} \ T^{\text{m}} + \dfrac{c^{\text{m}}}{12} \left(6 (\partial \tilde \gamma)^2 \tilde \gamma^{-4} - 3 (\partial^2 \tilde \gamma) \tilde \gamma^{-3} \right) \right)_{-2} \ket{j,m}^{(1)}\\
& = L^{\text{m}}_{-2} \ket{j,m+2}^{(1)} + \dfrac{c^{\text{m}}}{2} \tilde{\gamma}_{-1}\tilde{\gamma}_{-1} \ket{j,m+4}^{(1)} - \dfrac{c^{\text{m}}}{2} \tilde{\gamma}_{-2} \ket{j,m+3}^{(1)} \ , 
\end{align}
where we have used that $(\tilde \gamma^{-n})_0 \ket{j,m}^{(1)} =\ket{j, m+n}^{(1)}$, see Appendix~\ref{subapp:prop-gamma}. 

We can similarly compute the remaining terms on the left hand side by inserting the definitions \eqref{eq:spacetime-Vir} and \eqref{eq:matter Virasoro spacetime}, and using the Thielemans OPE package \cite{Thielemans:1991uw}; this leads to 
\begin{align}
\left( \LL_{-2} + b^{\pm 2} (\LL_{-1})^2 \right) \ket{j,m}^{(1)} &= l_{++}\tilde J_{-1}^+ \tilde J_{-1}^+ \ket{j,m}^{(1)} + l_{3+} \tilde J_{-1}^3 \tilde J_{-1}^+ \ket{j,m+1}^{(1)} \nonumber\\
 & \qquad + l_{-+} \tilde J_{-1}^- \tilde J_{-1}^+ \ket{j,m+2}^{(1)} + l_{-3} \tilde J_{-1}^- \tilde J_{-1}^3 \ket{j,m+3}^{(1)} \nonumber\\
 & \qquad + l_{--} \tilde J_{-1}^- \tilde J_{-1}^- \ket{j,m+4}^{(1)} + l_{33} \tilde J_{-1}^3 \tilde J_{-1}^3 \ket{j,m+2}^{(1)} \nonumber\\
& \qquad + l_3 \tilde J_{-2}^3 \ket{j,m+2}^{(1)} + l_+ \tilde J_{-2}^+ \ket{j,m+1}^{(1)} \nonumber\\
& \qquad + l_- \tilde J_{-2}^- \ket{j,m+3}^{(1)} - L_{-2}^{\text{m}} \ket{j,m+2}^{(1)} \ ,  \label{eq:LHS-ansatz}
\end{align}
where the coefficents $l_{++}$, $l_{3+}$, $l_{-+}$, $l_{-3}$, $l_{--}$, $l_{33}$, $l_3$, $l_+$, $l_-$, are given explicitly in Appendix~\ref{app:LHS-RHS-coeff}. Here we have expressed $c^{\text{m}}$ in terms of $k$ as in eq.~(\ref{cm}).

\paragraph{Rewriting as BRST exact state}

Next we want to rewrite (\ref{eq:LHS-ansatz}) in terms of the ansatz (\ref{eq:thesis-lev-two}). We first note that the ${L_{-2}^{\text{m}}}$ term in (\ref{eq:LHS-ansatz}) can only come from $L_{-2} = L_{-2}^{\text{m}} + L_{-2}^{\mathfrak{sl}(2, \mathds{R})}$ on the right-hand-side; thus the first term of the right-hand-side has to be 
\be 
\left( \LL_{-2} + b^{\pm 2} (\LL_{-1})^2 \right) \ket{j,m}^{(1)} = - \left(L_{-2} + \dfrac{3}{2} L_{-1}^2 \right) \ket{j,m+2}^{(1)} + L_{-1} \ket{\chi_1} \ .
\label{eq:LHS=RHS}
\ee 
Using the Sugawara representation, the right-hand-side of eq.~\eqref{eq:LHS=RHS} can now be written in terms of affine descendants, and we find 
\begin{align}
 &-\left(L_{-2} + \dfrac{3}{2} L_{-1}^2 \right) \ket{j,m+2}^{(1)} + L_{-1} \ket{\chi_1}  \nonumber\\
&\ \, =  r_3 \tilde J_{-2}^3 \ket{j,m+2}^{(1)} + r_+ \tilde J_{-2}^+ \ket{j,m+1}^{(1)}+ r_- \tilde J_{-2}^- \ket{j,m+3}^{(1)} -L_{-2}^{\text{m}} \ket{j,m+2}^{(1)}\nonumber\\
&\ \, \qquad  + r_{-3} \tilde J_{-1}^- \tilde J_{-1}^3 \ket{j,m+3}^{(1)} + r_{--} \tilde J_{-1}^- \tilde J_{-1}^- \ket{j,m+4}^{(1)} + r_{33} \tilde J_{-1}^3 \tilde J_{-1}^3 \ket{j,m+2}^{(1)}\nonumber\\
&\ \, \qquad +r_{++} \tilde J_{-1}^+ \tilde J_{-1}^+ \ket{j,m}^{(1)} + r_{3+} \tilde J_{-1}^3 \tilde J_{-1}^+ \ket{j,m+1}^{(1)} + r_{-+} \tilde J_{-1}^- \tilde J_{-1}^+ \ket{j,m+2}^{(1)} \ , 
\label{eq:RHS-ansatz} 
\end{align}
where the coefficents $r_{++}$, $r_{3+}$, $r_{-+}$, $r_{-3}$, $r_{--}$, $r_{33}$, $r_3$, $r_+$ and $r_-$ are spelled out explicitly in Appendix~\ref{app:LHS-RHS-coeff}. We therefore have to solve the system of equations 
\be
l_{ab}  = r_{ab} \ , \qquad
l_a = r_a \ , \qquad a,\, b \in \{\pm,\, 3\}\ .
\ee
Quite remarkably, this highly overconstrained system has a solution; in fact, there are two solutions
\begin{subequations}
\begin{align}
b^2 & = \left( \frac{k-3\pm \sqrt{k^2-10k+17}}{2\sqrt{k-2}} \right)^2 \ , \label{eq:b^2}\\
x_1 & = \frac{1 \mp \sqrt{k^2-10 k+17}}{4(k-2)^2} \ , \\
x_2 & = \frac{43 -86k +49 k^2 -8 k^3 \mp (8k^2-63 k + 115) \sqrt{k^2-10 k+17}}{16 (k-2)^2} \ , 
\end{align}
\end{subequations}
that are related to one another under the symmetry $b \to b^{-1}$. In addition, $j$ and $m$ must satisfy the conditions 
eq.~\eqref{eq:mass-shell-lev-two}. For each solution, the central charge of Liouville theory turns out to be 
\be 
c^\text{L} = 1 + 6 \bigl(b + b^{-1} \bigr)^2 = 1 + \dfrac{6(k-3)^2}{k-2} \ , 
\ee
in accordance with \eqref{eq:b string}. This is therefore a highly non-trivial confirmation of the conjecture of \cite{Eberhardt:2019qcl}.

\section{Null vectors in twisted sectors}\label{sec:twisted}

In this section we explain how the analysis of the previous section can be generalised for the twisted sectors. In the $w^{\rm th}$ twisted sector the Virasoro generators $\mathcal{L}_{n}$ are fractionally moded with $n \in \frac{1}{w}\mathds{Z}$, and they satisfy the commutation relations
\be 
[\mathcal{L}_m,\mathcal{L}_n] = (m-n) \mathcal{L}_{m+n} + \frac{w \ c_{\text{seed}}}{12} m (m^2-1) \delta_{m+n,0} \ , 
\ee
where $c_{\text{seed}}$ is the central charge of the seed theory. Following the strategy explained in Section~\ref{sec:null-states-CFT} one can again search for null vectors. At level $\frac{1}{w}$ the null vector is simply the $\mathcal{L}_{-\frac{1}{w}}$ descendant of the twisted sector ground state $|\sigma_w\rangle$,
\be
|N_1\rangle^{(w)} = \mathcal{L}_{-\frac{1}{w}} \, |\sigma_w\rangle \ , 
\ee
where the ground state has conformal weight 
\be
h^{(w)} = \dfrac{c_{\text{seed}}}{24} \dfrac{(w^2 - 1)}{w}  \ . 
\ee
Indeed, this is compatible with the commutation relations in the $w^{\rm th}$ twisted sector since we have 
\be 
0 = \bigl[ \mathcal{L}_{\frac{1}{w}}, \mathcal{L}_{-\frac{1}{w}} \bigr]  \ket{\sigma_w} = \frac{2}{w} \Bigl( h^{(w)} - \frac{c_{\text{seed}}}{24} \frac{(w^2 - 1)}{w} \Bigr) \ket{\sigma_w} \ . 
\ee
At level $2$ the null vector turns out to be 
\be 
\ket{N_{2}}^{(w)} =  \bigl( \mathcal{L}_{-\frac{2}{w}} + b^{\pm 2}  \mathcal{L}_{-\frac{1}{w}}^2 \bigr) \ket{h_{2}^{(w)}} \ , 
\ee
provided that 
\be
h_2^{(w)} = - \Bigl( \frac{1}{2} + \frac{3}{4} b^{\mp 2} \Bigr) + \frac{c_{\text{seed}}}{24} \frac{(w^2-1)}{w} \ . 
\ee
Note that these equations reduce to what we have seen above, see in particular eq.~(\ref{3.7}), for the case $w=1$. 

\subsection{Level one from the worldsheet}

Following the logic of Section~\ref{sec:CFTnull=BRSTexact}, we want to show that $|N_1\rangle^{(w)}$ is equal to an $L_{-1}$ descendant on the worldsheet. We start by identifying the $w$-twisted ground state with a worldsheet primary from the $w^{\rm th}$ spectrally flowed sector,
\be 
\ket{\sigma_w} = \ket{j,m}^{(w)}  \ . 
\label{eq:twisted-ground-state} 
\ee  
The $\mathcal{L}_0$ eigenvalue as well as the mass-shell condition impose 
\be 
m + \dfrac{k w}{2} = \dfrac{k (w^2 - 1)}{4w} \ , \qquad - \frac{j(j-1)}{k-2} - wm - \dfrac{k}{4} w^2 = 1 \ , 
\ee
where we have used that $c_{\text{seed}}= 6 k$, and thus $m =  -\frac{k(w^2+1)}{4w}$, while $j$ is, as before, see eq.~(\ref{3.14}), either 
\be\label{jsol}
j  = \frac{k}{2}-1  \ , \qquad \hbox{or} \qquad  j = 2 - \frac{k}{2} \ . 
\ee
The calculation of $\mathcal{L}_{-\frac{1}{w}} \, |\sigma_w\rangle$ is spelled out in Appendix~\ref{app:generic-w}, see eq.~(\ref{eq:-1/wstate}). This expression should now be equal to 
\be\label{eq:w-flowed-BRST}
\LL_{-\frac{1}{w}}\ket{\sigma_w} = \alpha\,   L_{-1} \ket{j,m+\tfrac{1}{w}}^{(w)}  \ , 
\ee
where $\alpha$ is some constant, and the right-hand-side can be evaluated using the Sugawara construction for the worldsheet Virasoro algebra; this leads to  
\begin{align}
L_{-1} \Ket{j,m+\tfrac{1}{w}}^{(w)} = & \frac{(wj+wm+ 1)}{w(k-2)}\tilde J_{-1}^- \Ket{j, m + \tfrac{1}{w}+1}^{(w)} \\
& -\frac{2(mw+1)+w^2(k-2)}{w(k-2)} \tilde J_{-1}^3 \Ket{j, m + \tfrac{1}{w}}^{(w)} \\
& + \frac{(1-wj+wm)}{w(k-2)}  \tilde J_{-1}^+ \Ket{j, m + \tfrac{1}{w}-1}^{(w)} \ .
\end{align}
Comparing with eq.~(\ref{eq:-1/wstate}), we therefore need to solve the equations 
\begin{subequations}
\begin{align}
\frac{(w-1) (-2 j+k w+2)}{2 w^2 (2 j+k-2)} & = \alpha \frac{(wj+wm+ 1)}{w(k-2)} \ , \\
\frac{2 \left(w^2-1\right) (j-1)}{w^2 (2 j+k-2)} & = -\alpha \frac{2(mw+1) +w^2(k-2)}{w(k-2)} \ , \\
\frac{(w+1) (2 j+k w-2)}{2 w^2 (2 j+k-2)} & = \alpha \dfrac{(1-wj+wm)}{w(k-2)} \ ,  
\end{align}
\end{subequations} 
and they have in fact the simple solution 
\be 
\alpha = -\dfrac{1}{w} \ , \qquad m = -\dfrac{k(w^2+1)}{4w} \ , \qquad j = \dfrac{k}{2} - 1 \ . 
\ee
Note that the equations for $m$ and $j$ are indeed compatible with (the first solution of) eq.~(\ref{jsol}). 

\subsection{The level two null vector} 

For the level $\frac{2}{w}$ null vector $\ket{N_{2}}^{(w)}$, we need to show that 
\be \label{eq:BRST-exact-level-2/w}
\left( \LL_{-\frac{2}{w}} + b^{\pm 2} \left( \LL_{-\frac{1}{w}} \right)^2 \right) \ket{j,m}^{(w)} = \alpha \left(L_{-2} + \dfrac{3}{2} L_{-1}^2 \right) \Ket{j,m + \tfrac{2}{w}}^{(w)} + L_{-1} \ket{\chi_1} \ . 
\ee
Here $\Ket{j,m}^{(w)}$ is the image of the affine primary state in the $w^{\rm th}$ twisted sector that corresponds to $\ket{h_{2}^{(w)}}$, i.e.\ $j$ and $m$ are characterised by 
\be 
h_{2}^{(w)} = m + \frac{wk}{2} \ , \qquad -\dfrac{j(j-1)}{k-2} - m w - \dfrac{k w^2}{4} = 1 \ . 
\label{eq:m-j-level-2/w}
\ee
The ansatz for the Virasoro primary state $\ket{\chi_1}$ is 
\begin{align}
\ket{\chi_1} &= \left(x_2\left(m+\frac{2}{w}+j-1\right)(w-1)-x_1\left(m+\frac{2}{w}-j+1\right)(w+1)\right) \tilde{J}^3_{-1} \Ket{j,m+\tfrac{2}{w}}^{(w)} \nonumber\\
&\quad\ \, +\left(m+\frac{2}{w}+\frac{kw}{2}\right)\left(x_1 \tilde{J}^-_{-1} \Ket{j,m+1+\tfrac{2}{w}}^{(w)}+x_2 \tilde{J}^+_{-1} \Ket{j,m-1+\tfrac{2}{w}}^{(w)}\right)\ ,
\end{align}
where $x_1$ and $x_2$ are arbitrary parameters. The left-hand-side of \eqref{eq:BRST-exact-level-2/w} can be computed along the lines of Section~\ref{sec:level-two-BRST-exact} by taking OPEs of DDF operators with vertex operators, while we simply employ the Sugawara construction to evaluate the right-hand-side. Demanding eq.~\eqref{eq:BRST-exact-level-2/w} then amounts to solving a system of nine equations in the six unknowns $\alpha$, $b$, $x_1$, $x_2$, $j$, and $m$. The two solutions we find have $j$ and $m$ satisfying eq.~\eqref{eq:m-j-level-2/w} and $b^2$ satisfying eq.~\eqref{eq:b^2}, corresponding again to a central charge 
\be 
c^\text{L} = 1 + 6 \bigl(b + b^{-1} \bigr)^2 = 1 + \dfrac{6(k-3)^2}{k-2} \ , 
\ee
in perfect agreement with the prediction of \cite{Eberhardt:2019qcl}.

\section{Constraining the correlators}\label{sec:BPZ-from-the-worldsheet}

\subsection{From null vectors to correlators}

The null vectors of the symmetric orbifold of Liouville theory that we have studied in the previous two sections constrain the correlation functions of the dual CFT. For example, the level two null vector (\ref{N2}) from the untwisted sector leads to the constraint that the 
$n$-point correlation functions of the untwisted sector Liouville theory have to satisfy 
\be 
\left\langle \Bigl[ \bigl( \mathcal{L}_{-2}^{\mathrm{L}}  + b^{\pm 2} \big(\mathcal{L}_{-1}^{\mathrm{L}}\big)^2\bigr) V_{h_*}\Bigr] (x_1) \, V_{h_2}(x_2) \cdots V_{h_n}(x_n) \right\rangle=0\ .
\label{eq:null-relation}
\ee
Using usual CFT arguments, this can be turned into a differential equation for the corresponding correlator, and one finds the BPZ differential equation \cite{Belavin:1984vu}
\begin{align}
\left( \sum_{i=2}^n \left(\frac{h_i}{(x_1-x_i)^2}+\frac{\partial_{x_i}}{x_1-x_i}\right)+ b^{\pm 2} \partial_{x_1}^2 \right) \left\langle V_{h_*}(x_1) V_{h_2}(x_2) \cdots V_{h_n}(x_n) \right\rangle=0\ .\label{eq:BPZ from dual CFT}
\end{align}
Following  \cite{Teschner:1995yf}, see also \cite{Ribault:2014hia} for a pedagogic introduction to the subject, one can use this BPZ equation together with crossing symmetry, to derive a shift equation for the structure constants of the theory. This shift equation in turn has a unique solution that reproduces the famous DOZZ formula \cite{Dorn:1992at, Dorn:1994xn, Zamolodchikov:1995aa}. 
Thus the knowledge of the level two null vector (\ref{N2}) is sufficient to show that the relevant theory is indeed Liouville theory.

One may hope that also the other null vectors (in particular, those that arise from the twisted sectors) will imply that the full theory is indeed the symmetric orbifold of Liouville theory, but we have not succeeded with this programme so far. One bottleneck is that we have not managed to turn the null vectors from the twisted sectors into simple differential equations. This is a generic problem with symmetric product theories whose solution does not seem to be known. 
\medskip

From what we have explained above, it should be clear that these differential equations should also be visible directly from a worldsheet perspective. Indeed, the corresponding null vectors are BRST exact on the worldsheet (and hence vanish inside physical correlation functions), and the only issue is whether they lead indeed to the correct  differential equation (with respect to the spacetime $x_i$ variables) as the BPZ equation (\ref{eq:BPZ from dual CFT}); this will be shown momentarily (see Section~\ref{sec:BPZworld}). Note that, using the logic of the above argument, this then allows us to deduce that the untwisted sector of the spacetime CFT contains indeed Liouville theory.

\subsection{BPZ equation from the worldsheet}\label{sec:BPZworld}
Let us recall that the string worldsheet theory based on $\mathfrak{sl}(2,\mathds{R})_k$ contains a BRST exact vector in the first spectrally flowed sector of the form
\be 
\Bigl[\bigl(\mathcal{L}_{-2}^{\mathrm{L}}  + b^{\pm 2} \big(\mathcal{L}_{-1}^{\mathrm{L}}\big)^2  \bigr) V_{h_*} \Bigr] (x;z)\ ,
\ee
where $h_*$ is the conformal weight (in the dual CFT). Thus, the worldsheet correlator vanishes
\be 
\left\langle\Bigl[\bigl(\mathcal{L}_{-2}^{\mathrm{L}}  + b^{\pm 2} \big(\mathcal{L}_{-1}^{\mathrm{L}}\big)^2  \bigr) V_{h_1} \Bigr](x_1;z_1)  \, V_{h_2}(x_2;z_2) \cdots V_{h_n}(x_n;z_n) \right\rangle=0\ ,\label{eq:null-relation-ws}
\ee
where we have set $h_1 = h_*$, and we have assumed that all vertex operators are physical, i.e.\ in particular Virasoro primary. 
Mirroring the untwisted sector of the symmetric product orbifold, we shall assume that all vertex operators are in the $w=1$ spectrally flowed sector. We will comment on generic spectral flow in Section~\ref{sec:conclusions}. 

We now want to rederive the differential equation \eqref{eq:BPZ from dual CFT} from this perspective, i.e.\ starting from \eqref{eq:null-relation-ws}. We rewrite \eqref{eq:null-relation-ws} as the difference of the total and the matter CFT contribution,  
\begin{multline}
\Bigl\langle \Bigl[\bigl(\mathcal{L}_{-2} + b^{\pm 2} \ \mathcal{L}_{-1}^2 \bigr) V_{h_1}\Bigr](x_1, z_1)\,  V_{h_2}(x_2, z_2) \cdots V_{h_n}(x_n, z_n) \Bigr\rangle  \\
\ \  +\Bigl\langle \Bigl[\bigl(-\mathcal{L}_{-2}^{\mathrm{m}}  - 2 b^{\pm 2}\mathcal{L}_{-1} \LLm_{-1}  + b^{\pm 2} \big(\mathcal{L}_{-1}^{\mathrm{m}}\big)^2 \bigr)V_{h_1}\Bigr](x_1, z_1) V_{h_2}(x_2, z_2) \cdots V_{h_n}(x_n, z_n) \Bigr\rangle =0 \ , \label{eq:null-relation-ws-tot-m}
\end{multline}
where we have used that $[\mathcal{L}_{-1}, \LLm_{-1}] = 0 $. Let us start from the first line in \eqref{eq:null-relation-ws-tot-m}. The $\mathcal{L}_{-1}^2$ piece is easy to evaluate and gives immediately the same result as in the calculation above,
\be 
\bigl(\mathcal{L}_{-1}^2 V_{h_1} \bigr) (x;z) = \bigl( (J_0^+)^2 V_{h_1} \bigr) (x;z) =   \partial_x^2 V_{h_1}(x;z)  \ .
\ee 
Next we evaluate the other term from the first line, i.e.\ the term containing $\mathcal{L}_{-2}$, 
\begin{align}
& \left\langle (\mathcal{L}_{-2}V_{h_1})(x_1, z_1) V_{h_2}(x_2, z_2) \cdots V_{h_n} (x_n, z_n) \right\rangle \nonumber \\
& \qquad  = \left\langle (\mathcal{L}_{-2}V_{h_1})(0, z_1) V_{h_2}(x_2-x_1, z_2) \cdots V_{h_n} (x_n-x_1, z_n) \right\rangle \\
& \qquad  = \left\langle \oint_{z_1} \mathrm{d}\zeta \ \left(- \gamma^{-2} J^3 + 2 \gamma^{-1} J^+ \right)(\zeta)V_{h_1}(0;z_1) \prod_{j \neq 1} V_{h_j}(x_j-x_1, z_j) \right\rangle \\
& \qquad  = - \left\langle V_{h_1}(0;z_1) \sum_{i = 2}^n \oint_{z_i} \mathrm{d}\zeta \ \left(- \gamma^{-2} J^3 + 2 \gamma^{-1} J^+ \right)(\zeta) \prod_{j \neq 1} V_{h_j}(x_j-x_1, z_j) \right\rangle \\
& \qquad  = - \Bigl\langle \sum_{i = 2}^n V_{h_1}(x_1-x_i;z_1) \oint_{z_i} \mathrm{d}\zeta \ \mathrm{e}^{(x_1-x_i) J_0^+} \left(- \gamma^{-2} J^3 + 2 \gamma^{-1} J^+ \right)(\zeta) \, \mathrm{e}^{-(x_1-x_i) J_0^+}  \nonumber \\
& \qquad \qquad \qquad \qquad \qquad \qquad \qquad  \times V_{h_i}(0, z_i) \prod_{j \neq 1, j \neq i} V_{h_j}(x_j-x_i, z_j) \Bigr\rangle \ ,
\end{align}
where in the first step we have shifted all $x_j$ by $-x_1$, while in the last step we have shifted them by $x_1 - x_i$.  (Here we have used (\ref{doubletranslation}).) 
It follows from the commutators in Appendix~\ref{app:conventions} that
\begin{subequations}
\begin{align}
\mathrm{e}^{(x_1-x_i) J_0^+} \, \gamma(\zeta) \, \mathrm{e}^{-(x_1-x_i) J_0^+} &= \gamma(\zeta) - (x_1 - x_i) \ , \\ 
\mathrm{e}^{(x_1-x_i) J_0^+} \, J^3(\zeta) \, \mathrm{e}^{-(x_1-x_i)  J_0^+} &= J^3(\zeta) - (x_1-x_i)  \, J^+(\zeta) \ , 
\end{align}
\end{subequations}
and thus
\begin{align}
& \left\langle (\mathcal{L}_{-2}V_{h_1})(x_1, z_1) V_{h_2}(x_2, z_2) \cdots V_{h_n} (x_n, z_n) \right\rangle \\
& \qquad  =  \Biggl\langle \sum_{i = 2}^n V_{h_1}(x_1-x_i;z_1) \oint_{z_i} \mathrm{d}\zeta \ \left(  \dfrac{J^3}{(\gamma - x_1 + x_i)^2} \right)(\zeta) \, V_{h_i}(0, z_i) \prod_{j \neq 1, j \neq i} V_{h_j}(x_j-x_i, z_j)  \Biggr\rangle \nonumber\\
& \qquad  \qquad  +  \Biggl\langle  \sum_{i = 2}^n V_{h_1}(x_1-x_i;z_1) \oint_{z_i} \mathrm{d}\zeta \ \left( \dfrac{(x_1-x_i-2 \gamma)}{(\gamma - x_1 + x_i)^2} J^+ \right)(\zeta) \, V_{h_i}(0, z_i)  \nonumber\\
& \qquad \qquad \qquad \qquad \qquad \qquad \qquad  \qquad \qquad \qquad \times \prod_{j \neq 1, j \neq i} V_{h_j}(x_j-x_i, z_j) \Biggr\rangle \ .
\end{align} 
In order to proceed, we now evaluate (without loss of generality we may assume that $z_i=0$)
\begin{align}
& \oint_0 \mathrm{d}\zeta\ \left( \dfrac{(x_1 - x_i - 2 \gamma)}{(\gamma - x_1 + x_i)^2} J^+ \right)(\zeta) V_{h_i}(0,0) = \left[ \left( \dfrac{(x_1 - x_i - 2 \gamma)}{(\gamma - x_1 + x_i)^2} J^+ \right)_0 V_{h_i} \right](0,0)\\
& \quad = \sum_{n \leq 0} \left( \dfrac{(x_1 - x_i - 2 \gamma)}{(\gamma - x_1 + x_i)^2} \right)_n J^+_{-n} V_{h_i}(0,0) + \sum_{n \geq 1} J^+_{-n} \left( \dfrac{(x_1 - x_i - 2 \gamma)}{(\gamma - x_1 + x_i)^2} \right)_n V_{h_i}(0,0) \\
& \quad = \left( \dfrac{(x_1 - x_i - 2 \gamma)}{(\gamma - x_1 + x_i)^2} \right)_0 J^+_0 V_{h_i}(0,0) + \left( \dfrac{(x_1 - x_i - 2 \gamma)}{(\gamma - x_1 + x_i)^2} \right)_{-1} J^+_1 V_{h_i}(0,0) \\
& \quad = \dfrac{\partial_x V (0,0)}{x_1- x_i} + \left( \dfrac{x_1 - x_i - 2 \gamma}{(\gamma - x_1 + x_i)^2} \right)_{-1} J^+_1 V_{h_i}(0,0) = \dfrac{\partial_x V_{h_i} (0,0)}{x_1 - x_i} \ , 
\end{align}
where the last step follows from observing that 
\begin{align}
& \left( \dfrac{x_1 - x_i - 2 \gamma}{(\gamma - x_1 + x_i)^2} \right)_{-1} = \left( \partial \left( \dfrac{x_1 - x_i - 2 \gamma}{(\gamma - x_1 + x_i)^2} \right) \right)_{-1} = \left(\dfrac{2 \gamma \partial \gamma}{(\gamma - x_1 + x_i)^3} \right)_{-1} \\
& \qquad = 2 \left(\dfrac{\gamma}{(\gamma - x_1 + x_i)^3} \right)_{-1} (\partial \gamma)_0 + 2 \left(\dfrac{\gamma}{(\gamma - x_1 + x_i)^3} \right)_{0} (\partial \gamma)_{-1}  \ ,
\end{align}
together with $(\partial \gamma)_0 = 0$ and $\gamma_0 V_{h_i}(0,0) = 0 $. A similar, but easier, calculation yields
\begin{align}
& \oint_0 \mathrm{d}\zeta\  \left( \dfrac{J^3}{(\gamma - x_1 + x_i)^2}  \right)(\zeta) V_{h_i}(0,0) = \dfrac{h_i V_{h_i}(0,0)}{(x_1-x_i)^2}\ .
\end{align}
 Putting everything together we thus find 
\begin{multline}
\left\langle (\mathcal{L}_{-2}V_{h_1})(x_1, z_1) V_{h_2}(x_2, z_2) \cdots V_{h_n} (x_n, z_n) \right\rangle \\
  =  \sum_{i=2}^n \left( \dfrac{\partial_{x_i}}{x_1 - x_i} + \dfrac{h_i}{(x_1-x_i)^2} \right) \left\langle V_{h_1}(x_1, z_1) \cdots V_{h_n}(x_n, z_n) \right\rangle \ , 
\end{multline}
which therefore reproduces the rest of the BPZ equation. 
\smallskip

It only remains to show that the matter CFT, i.e.\ the second line in (\ref{eq:null-relation-ws-tot-m}), does not contribute. Obviously, the terms involving $(\LLm_{-1} V_{h_*})(x;z_1)=0$ vanish, since the vertex operator is in the vacuum state with respect to the matter CFT, see also the comment below eq.~(\ref{eq:mass-shell-lev-two}). As regards $(\LLm_{-2} V_{h_*})(0,0)$, this was already calculated in \eqref{eq:matter Virasoro level 2}, and was expressed in terms of $T^\text{m}$ and $\tilde\gamma$  descendants. Since both of these fields have regular OPEs with all $V_{h_i}(x_i;z_i)$, it follows that the contribution vanishes after performing the usual contour deformation argument. Thus we have shown that the second line in (\ref{eq:null-relation-ws-tot-m}) vanishes, and hence that \begin{align}
\left( \sum_{i=2}^n\left(\frac{h_i}{(x_1-x_i)^2}+\frac{ \partial_{x_i}}{x_1-x_i}\right) + b^{\pm 2} \partial_{x_1}^2 \right) \left\langle V_{h_1}(x_1;z_1)\cdots  V_{h_n}(x_n;z_n)\right \rangle=0\ .
\label{eq:BPZ-final}
\end{align}
This is essentially the same formula as eq.~(\ref{eq:BPZ from dual CFT}), except that the correlation function is now a worldsheet correlator. In string theory we have to perform the moduli space integral, i.e.\ fix 3 positions (say $z_1=0$, $z_2=1$ and $z_3=\infty$) and integrate over the remaining $n-3$ variables. The differential equation is however independent of the positions $z_i$, and hence (\ref{eq:BPZ-final}) implies indeed  (\ref{eq:BPZ from dual CFT}).

We have therefore shown that the sphere correlators with one degenerate field inserted satisfy the same differential equation both in the untwisted sector of the dual CFT and in the $w=1$ sector on the worldsheet.
Following the technology developed in \cite{Teschner:1995yf}, this allows one to deduce that the three-point functions agree on both sides of the duality. Matching three-point functions implies then also that all untwisted correlation functions (at least of single particle states) agree. This goes a long way towards proving the duality proposed in \cite{Eberhardt:2019qcl}.

\section{Conclusions}\label{sec:conclusions}

In this paper we have studied bosonic string theory on $\text{AdS}_3\times X$, and tested the conjectural duality with the two-dimensional CFT \cite{Eberhardt:2019qcl}
\be 
\text{Sym}^N\left(\left(\text{Liouville with $c^{\text L}=1+\frac{6(k-3)^2}{k-2}$} \right)\times X \right)\ . 
\label{eq:bosonic-dual-CFT}
\ee
We have shown that the low-lying null vectors of Liouville theory correspond precisely to BRST exact states in the worldsheet theory. As a consequence, the correlation functions on the two sides of the duality obey the same constraints. These constraints were enough to show that the seed theory of the symmetric product orbifold CFT contains indeed the Liouville factor, with the central charge as proposed in \cite{Eberhardt:2019qcl}. 

The matching of the null vector structure implies, in particular, that correlation functions of fields in the $w=1$ spectrally flowed sector of the worldsheet theory coincide with correlation functions of untwisted operators in the dual CFT. Among other things, this therefore determines certain spectrally flowed $\mathrm{SL}(2,\mathds{R})_k$ correlation functions that had not appeared in the literature before.
\smallskip

We have also generalised the analysis to the low-lying null vectors for generic $w$-twisted sectors, and we have shown that they also correspond to BRST exact states in the worldsheet description. While this implies that they vanish again in correlation functions, we have not managed to turn these null vector constraints into differential equations, neither on the worldsheet nor in the symmetric orbifold. This is essentially because the usual contour arguments do not apply directly to fractionally moded generators.\footnote{More conceptually, this probably also reflects that the symmetric orbifold correlators are controlled by the associated covering map \cite{Lunin:2000yv, Pakman:2009zz}, and that the covering space is typically of higher genus.} This seems to be a general problem of symmetric orbifold theories, and it would be nice to make progress in this direction. 

We have concentrated on the null vectors that appear in the dual CFT (in particular, in the Liouville factor), but one could also ask about the null vectors that exist in the worldsheet theory, i.e.\ in particular in the $\mathfrak{sl}(2,\mathds{R})_k$ factor. 
However, we have not managed to turn them into constraints on the spacetime correlators.
It would be interesting though to explore this also further.

In this paper we have studied bosonic string theory on $\text{AdS}_3$ together with its (bosonic) Liouville dual, see eq.~\eqref{eq:bosonic-dual-CFT}. We expect that our analysis can be generalised to supersymmetric incarnations of the duality, in particular to the maximally supersymmetric backgrounds $\text{AdS}_3 \times \text{S}^3 \times \mathbb{T}^4$ \cite{Eberhardt:2019qcl} and $\text{AdS}_3 \times \text{S}^3 \times \text{S}^3 \times \text{S}^1$ \cite{Eberhardt:2019niq}. For $\text{AdS}_3 \times \text{S}^3 \times \mathbb{T}^4$ we have checked that the lowest level null vector, namely the one corresponding to 
\be
\mathcal{G}_{-1/2}^{\alpha\beta}\ket{0} = 0 
\ee
in the dual CFT, is indeed BRST exact from the worldsheet perspective. (Here $\mathcal{G}_r^{\alpha\beta}$ are the supercharges of the dual $\mathcal{N}=4$ superconformal algebra.) It would be interesting to push this analysis to some higher null vectors. For bosonic Liouville theory, the null vector constraints (together with crossing symmetry) are sufficient to determine the structure constants uniquely \cite{Teschner:1995yf}, and this argument can be generalised to $\mathcal{N}=1$ Liouville theory \cite{Poghosian:1996dw}. However, this line of reasoning seems to fail for the case with $\mathcal{N} \ge 2$ supersymmetry, and other methods will have to be developed.

\section*{Acknowledgements}

We thank Rajesh Gopakumar, Elli Pomoni, Leonardo Rastelli, Alessandro Sfondrini, Yifan Wang and Xi Yin for useful conversations. LE \& MRG thank the Erwin Schr\"odinger Institute in Vienna  for hospitality during the thematic programme ``Higher Spins and Holography", while AD thanks the Galileo Galilei Institute for Theoretical Physics and INFN for hospitality during the workshop ``String Theory from a worldsheet perspective". The work of AD and LE is supported by the Swiss National Science Foundation, and all three of us acknowledge support by the NCCR SwissMAP, which is also funded by the Swiss National Science Foundation.  

\appendix 

\section{Conventions}\label{app:conventions}

\subsection[The \texorpdfstring{$\text{SL}(2,\mathds{R})_k$}{SL(2,R)k} WZW model and its Wakimoto representation]{The $\boldsymbol{\text{SL}(2,\mathds{R})_k}$ WZW model and its Wakimoto representation}

The $\mathfrak{sl}(2,\mathds{R})_k$ algebra is generated by the modes $J^a_n$ with $a=\pm, 3$, satisfying the commutation relations 
\begin{subequations}
\begin{align} 
[J^3_m,J^3_n]&=-\tfrac{1}{2}km\delta_{m+n,0}\ ,  \label{eq:sl2 commutation relations a}\\
[J^3_m,J^\pm_n]&=\pm J^\pm_{m+n}\ , \label{eq:sl2 commutation relations b}\\
[J^+_m,J^-_n]&=km\delta_{m+n,0}-2J^3_{m+n}\ . \label{eq:sl2 commutation relations c}
\end{align}
\end{subequations}
We shall also need its Wakimoto representation in terms of a pair of bosonic ghosts
\be 
\beta(z)\gamma(y) \sim - \dfrac{1}{z-y} \ , 
\ee
and a free boson 
\be 
\partial \Phi(z)\partial \Phi(y) \sim -\dfrac{1}{(z-y)^2} 
\ee
with background charge $Q=\sqrt{\frac{1}{k-2}}$. The $\mathfrak{sl}(2,\mathds{R})_k$ fields are then given as 
\begin{subequations}
\begin{align}
J^+ &= \beta \ , \\
J^3 &= \sqrt{\tfrac{k-2}{2}} \, \partial \Phi + (\beta \gamma) \ , \\
J^- &= \sqrt{2(k-2)}\, (\partial \Phi \gamma)+ (\beta \gamma \gamma)-k \partial \gamma \ . 
\end{align}
\label{eq:wakimoto}
\end{subequations}
The Wakimoto representation is a good approximation to the $\mathrm{SL}(2,\mathds{R})_k$ WZW model close to the boundary of $\mathrm{AdS}_3$ ($\Phi \to \infty$). To compute correlation functions in this representation, screening charges have to be included, see e.g.~\cite{Giribet:2000fy, Hosomichi:2000bm}.
We will not try to compute correlation functions directly in the Wakimoto representation, but will only use it when acting on states. 

Later on we will also need somewhat formal expressions of the Wakimoto fields, such as fractional powers of $\gamma$. We will compute them via analytic continuation, in the same spirit to what was done in \cite{Malikov:1986aa}. Some convenient formulae for this purpose are summarised in Appendix~\ref{subapp:prop-gamma}. 

\subsection{Vertex operators}

Affine highest weight representations of $\mathfrak{sl}(2,\mathds{R})_k$, together with their spectrally flowed images make up the spectrum of the WZW model. 
The unflowed ground state $\ket{j,m}^{(0)}$ can be expressed in terms of the Wakimoto representation as 
\be \label{jmvertex}
\ket{j,m}^{(0)} = \oint_0 \mathrm{d}z\  \gamma^{-j-m} e^{j \sqrt{\tfrac{2}{k-2}} \Phi}(z) \ket{0} \ .  
\ee

There are two kinds of highest weight representations that appear in the WZW spectrum. They correspond to $j \in \mathds{R}$ (the discrete representations) and $j=\tfrac{1}{2}+i\, \mathds{R}$ (the continuous representations), respectively. In the discrete case, we require in addition that $m-j \in \mathds{N}_0$, whereas in continuous representations, $m \in \mathds{Z}+\lambda$ is not related to $j$. In either case, the Casimir of the representation is given by
\be 
\mathcal{C}=-j(j-1)\ .
\ee
In the application to $\mathrm{AdS}_3$ string theory, discrete representations correspond to short strings, while continuous representations describe long string solutions \cite{Maldacena:2000hw}. 

\subsection{Spectral flow}

The spectral flow automorphism $\sigma$ acts on the modes of the affine algebra as\footnote{This is the spectral flow automorphism along the $J^3$ direction, and this is the one we shall be using on states.}
\begin{subequations}
\begin{align}
\sigma^w(J^\pm_m) & = J^{\pm}_{m\mp w}\ , \label{eq:spectral flow def a}\\
\sigma^w(J^3_m) & = J^3_m + \tfrac{kw}{2}\delta_{m,0} \ , \label{eq:spectral flow def b}
\end{align}
\end{subequations}
while the Sugawara Virasoro modes transform as 
\be 
\sigma^w(L^{\mathfrak{sl}(2,\mathds{R})}_m)=L^{\mathfrak{sl}(2,\mathds{R})}_m-w J^3_m-\frac{k}{4}w^2 \delta_{m,0}\ .
\ee
In terms of the Wakimoto representation, spectral flow therefore acts as 
\begin{subequations}
\begin{align}
\sigma^w(\gamma_m)&=\gamma_{m+w}\ , \\
\sigma^w(\beta_m)&=\beta_{m-w}\ , \\
\sigma^w(\partial\Phi_m)&=\partial \Phi_m+\sqrt{\tfrac{k-2}{2}} w \delta_{m,0} \ .
\end{align}
\end{subequations}
In the main text we have denoted modes before spectral flow with a tilde
\be
J_m^a = \sigma^w(\tilde J_m^a) \ , \qquad L_m^a = \sigma^w(\tilde L_m^a) \ .
\label{eq:tilde-def}
\ee
The (tilde) affine modes thus always act as in usual highest weight representations, i.e.\ 
\begin{subequations} 
\begin{align}
\tilde J_0^+ \ket{j,m}^{(w)} &= (m+j) \ket{j,m +1 }^{(w)} \ ,& \tilde J_n^+ \ket{j,m}^{(w)} &=0\ , & n&>0 \ , \\
\tilde J_0^3 \ket{j,m}^{(w)} &= m  \ket{j,m}^{(w)} \ , & \tilde J_n^3 \ket{j,m}^{(w)} &=0\ , & n&>0 \ , \\ 
\tilde J_0^- \ket{j,m}^{(w)} &= (m-j) \ket{j,m-1}^{(w)} \ , & \tilde J_n^- \ket{j,m}^{(w)} &=0\ , & n&>0 \ .
\end{align}
\label{eq:tilded-zero-modes}
\end{subequations}
In terms of the spectrally flowed modes, this then means 
\begin{subequations} 
\begin{align}
J_{w}^+ \ket{j,m}^{(w)} &= (m+j) \ket{j,m +1 }^{(w)} \ ,& J_n^+ \ket{j,m}^{(w)} &=0\ , & n&>w \ , \\
J_0^3 \ket{j,m}^{(w)} &= \left(m + \dfrac{k w }{2} \right) \ket{j,m}^{(w)} \ , & J_n^3 \ket{j,m}^{(w)} &=0\ , & n&>0 \ , \\ 
J_{-w}^- \ket{j,m}^{(w)} &= (m-j) \ket{j,m-1}^{(w)} \ , & J_n^- \ket{j,m}^{(w)} &=0\ , & n&>-w \ .
\end{align}
\end{subequations}

\subsection[Evaluating  \texorpdfstring{$\gamma$}{gamma}]{Evaluating $\boldsymbol{\gamma}$} \label{subapp:prop-gamma}

We also need to deduce some rules for how to evaluate expressions involving the Wakimoto field $\gamma$. First we note that it follows from \eqref{eq:wakimoto} as well as (\ref{jmvertex}), that
\be 
\tilde \gamma (z) \ket{j,m}^{(w)} = \ket{j,m-1}^{(w)} + z  \tilde \gamma_{-1} \ket{j,m}^{(w)} + \cdots \ , 
\label{eq:tilde-gamma-V-OPE}
\ee
where we have used that $\gamma_0 \ket{j,m} = \ket{j,m-1}$. 
By repeatedly applying \eqref{eq:tilde-gamma-V-OPE}, we obtain for $n \in \mathds{N}$ 
\be
\tilde \gamma^n(z) \ket{j,m}^{(w)}  = \ket{j,m-n}^{(w)} + n \, z \, \tilde \gamma_{-1} \ket{j,m-n+1}^{(w)} + \cdots \ .
\label{eq:tilde-gamman-V-OPE}
\ee
We will frequently analytically continue \eqref{eq:tilde-gamman-V-OPE} to $n \in \mathds{Q}$. Since we can rewrite eq.~\eqref{eq:tilde-gamman-V-OPE} as 
\be
\tilde \gamma^n(z) \ket{j,m}^{(w)} = \left(\tilde \gamma^n \right)_0 \ket{j,m}^{(w)} + z \, \left( \tilde \gamma^n \right)_{-1} \ket{j,m}^{(w)} + \cdots \ , 
\ee
we deduce by comparing terms that 
\be 
 \left(\tilde \gamma^n \right)_0 \ket{j,m} = \ket{j,m-n} \ , \qquad \left( \tilde \gamma^n \right)_{-1} \ket{j,m} = n \, \tilde \gamma _{-1} \ket{j,m-n+1} \ . 
\label{eq:gamma^n_0}
\ee

We also want to express the action of the modes of $\gamma$ on worldsheet primary vertex operators in terms of the affine Kac-Moody currents. In order to do so, we may invert the Wakimoto representation
\be 
\partial \gamma = -\dfrac{1}{(k-2)} \left( J^+ \gamma^2 - 2 J^3 \gamma + J^- \right) \ , 
\ee
and obtain 
\begin{align}
\tilde \gamma_{-1} \ket{j,m}^{(w)} &= -\dfrac{1}{(k-2)} \left( \tilde{J}^+ \tilde{\gamma}^2 - 2 \tilde{J}^3 \tilde{\gamma} + \tilde{J}^- \right)_{-1} \ket{j,m}^{(w)} \\
&= - \dfrac{1}{(k-2)} \left( \tilde J_{-1}^+ \ket{j,m-2}^{(w)} - 2 \tilde J_{-1}^3 \ket{j,m-1}^{(w)}  \right. \nonumber\\
& \qquad\quad\left.  + 2(j+m) \tilde \gamma_{-1} \ket{j,m}^{(w)} -2m \tilde \gamma_{-1} \ket{j,m}^{(w)} + \tilde J_{-1}^- \ket{j,m}^{(w)} \right) \ . 
\end{align}
Finally, we can solve for $\tilde \gamma_{-1} \ket{j,m}$ and thus deduce 
\be 
\tilde \gamma_{-1} \ket{j,m}^{(w)} = - \dfrac{1}{k-2+2j} \Bigl( \tilde J_{-1}^+ \ket{j,m-2}^{(w)} -2 \tilde J_{-1}^3 \ket{j,m-1}^{(w)} + \tilde J_{-1}^- \ket{j,m}^{(w)} \Bigr) \ . 
\label{eq:gamma-1}
\ee
The last ingredient we need are the OPEs 
\begin{subequations} 
\begin{align}
\tilde J^+(z) \tilde \gamma (\zeta) & = -\dfrac{1}{z-\zeta} + \cdots  \ , \\ 
\tilde J^3(z) \tilde \gamma (\zeta) & = -\dfrac{\tilde \gamma(\zeta)}{z-\zeta} + \cdots \ , \\ 
\tilde J^-(z) \tilde \gamma (\zeta) & = -\dfrac{(\tilde \gamma \tilde \gamma)(\zeta)}{z-\zeta} + \cdots \ , 
\end{align}
\end{subequations}
that follow directly from (\ref{eq:wakimoto}). 

\section{Coefficients for level two null state}
\label{app:LHS-RHS-coeff}

The coefficients $l_{++}$, $l_{3+}$, $l_{-+}$, $l_{-3}$, $l_{--}$, $l_{33}$, $l_3$, $l_+$ and $l_-$ entering eq.~\eqref{eq:LHS-ansatz} are
\begin{subequations}
\begin{align}
l_{++} = & \ C \ \Bigl( 12 j^2 k-24 j^2+24 j k^2-61 j k+20j \nonumber \\
& \qquad \quad  +6 k^3-6 k^2 m-42 k^2-11 k m+61 k+52 m+4 \Bigr) +b^{\pm 2}\ ,  \\
l_{3+} = & \ -2 C \ \Bigl(24 j^2 k-48 j^2+36 j k^2-144 j k+144j \ , \nonumber\\
 & \qquad \qquad \ +6 k^3-12 k^2 m-89 k^2-22 k m+185 k+104 m-44 \Bigr) \ , \\
l_{-+} = & \ 2 C \ \Bigl(12 j^2 k-24 j^2+12 j k^2-83 j k+124 j \nonumber\\
& \qquad \qquad -6 k^2 m-47 k^2-11 k m+124 k+52 m-48 \Bigr) \ , \\
l_{-3} =  & \ -2 C \ \Bigl(24 j^2 k-48 j^2+12 j k^2-188 j k \nonumber\\
& \qquad \qquad +352 j-6 k^3-12 k^2 m-99 k^2-22 k m+311 k+104 m-148 \Bigr) \ ,  \\
l_{--} = & \ C \ \Bigl(12 j^2 k-24 j^2-105 j k+228 j-6 k^3 \nonumber\\  
& \qquad \qquad -6 k^2 m -52 k^2-11 k m+187 k+52 m-100 \Bigr) \ , \\
l_{33} = & \ 4 C  \ \Bigl(12 j^2 k-24 j^2+12 j k^2-83 j k+124 j \nonumber\\ 
& \qquad \qquad -6 k^2 m-47 k^2-11 k m+124 k+52 m-48 \Bigr) \ , \\
l_3 = & \ -2 C \ \Bigl(24 j^3 k-48 j^3+24 j^2 k^2-212 j^2 k+352 j^2 \nonumber\\
& \qquad \qquad  + 6 j k^3 -146 j k^2 +501 j k -460 j-29 k^3 \nonumber\\
& \qquad \qquad -6 k^2 m  +139 k^2-11 k m-237 k+52 m+156 \Bigr) \ , \\
l_+ = & \ C \ \Bigl(48 j^3 k-96 j^3+72 j^2 k^2-380 j^2 k+496 j^2 \nonumber\\
& \qquad \qquad +36 j k^3 -284 j k^2+728 j k-608 j+6 k^4 \nonumber\\
& \qquad \qquad -65 k^3 -12 k^2 m +205 k^2-22 k m-296 k+104 m+208 \Bigr) \ , \\
l_- = & \ C \ \Bigl(16 j^3 k-32 j^3+8 j^2 k^2-140 j^2 k+272 j^2 \nonumber\\
& \qquad \qquad -4 j k^3-92 j k^2 +300 j k-176 j-2 k^4 \nonumber\\
& \qquad \qquad -21 k^3 -12 k^2 m+51 k^2-22 k m-10 k+104 m+40 \Bigr) \ , 
\end{align}
\end{subequations}
where the overall coefficient $C$ equals 
\be 
C =  \ \frac{1}{4 (k-2) (2 j+k-2) (j+k-2) (2 j+k-1)} \ .
\ee
\medskip

On the other hand, the coefficients $r_{++}$, $r_{3+}$, $r_{-+}$, $r_{-3}$, $r_{--}$, $r_{33}$, $r_3$, $r_+$ and $r_-$ entering eq.~\eqref{eq:RHS-ansatz} are
\begin{subequations} 
\begin{align}
r_{++} & = \frac{(j-m-1) (-3 j-2 k x_2 +3 m+4 x_2 +6)}{2 (k-2)^2} \ , \\
r_{3+} & = \frac{1}{(k-2)^2}\Bigl(-4 j^2 k x_1 +8 j^2 x_1 +8 j k m x_1 +20 j k x_1 -3 j k \nonumber\\
& \qquad \qquad \qquad -16 j m x_1 -6 j m -40 j x_1 -3 j -k^2 x_2 -4 k m^2 x_1  \nonumber\\
& \qquad \qquad \qquad -20 k m x_1 -2 k m x_2 +3 k m -24 k x_1 +2 k x_2 +6 k \nonumber\\
& \qquad \qquad \qquad +8 m^2 x_1 +6 m^2+40 m x_1 +4 m x_2 +15 m+48 x_1 +6 \Bigr) \ , \\
r_{-+} & = \frac{1}{(k-2)^2} \Bigl(3 j^2-j k^2 x_1 -2 j k m x_1 -2 j k x_1 +j k x_2 +4 j m x_1  \nonumber\\
& \qquad \qquad \qquad +8 j x_1 -2 j x_2 -3 j+k^2 m x_1 +3 k^2 x_1 +2 k m^2 x_1 \nonumber \\
& \qquad \qquad \qquad +8 k m x_1 +k m x_2 +6 k x_1 +k x_2 -k -4 m^2 x_1  \nonumber\\
& \qquad \qquad \qquad -3 m^2 -20 m x_1 -2 m x_2 -12 m -24 x_1 -2 x_2 -10 \Bigr) \ , \\
r_{-3} & = \frac{1}{{(k-2)^2}} \Bigl( 4 j^2 k x_1 -8 j^2 x_1 -4 j k x_1 +3 j k+6 j m \nonumber\\
& \qquad \qquad \qquad +8 j x_1 +9 j -k^3 x_1 -4 k^2 m x_1 -6 k^2 x_1  \nonumber\\
& \qquad \qquad \qquad -8 k m^2 x_1 -28 k m x_1 +3 k m -24 k x_1 +6 k \nonumber\\
& \qquad \qquad \qquad +16 m^2 x_1 +6 m^2+72 m x_1 +21 m+80 x_1 +18 \Bigr) \ , \\
r_{--} & = -\frac{(j+m+3)}{2(k-2)^2}\Bigr(3 j-2 k^2 x_1 -4 k m x_1 -4 k x_1 +8 m x_1 +3 m+16 x_1 +6 \Bigr)  \ , \\
r_{33} & = -\frac{1}{2 (k-2)^2}\Bigl(8 j k^2 x_1 +16 j k m x_1 -32 j m x_1 -32 j x_1 -8 k^2 m x_1\nonumber\\
& \qquad \qquad \qquad \quad  -24 k^2 x_1 +3 k^2 -16 k m^2 x_1 -48 k m x_1 +12 k m \nonumber\\
& \qquad \qquad \qquad \quad +10 k +32 m^2 x_1 +12 m^2+128 m x_1 +24 m+96 x_1 +16 \Bigr) \ , \\
r_3 & = \frac{1}{2 (k-2)^2} \Bigl(-6 j^2+8 j k^2 x_1 -32 j k x_1 -4 j k x_2  \nonumber\\
& \qquad \qquad \qquad +32 j x_1 +8 j x_2 +6 j-8 k^2 m x_1 -24 k^2 x_1 \nonumber \\
& \qquad \qquad \qquad +5 k^2 +32 k m x_1 -4 k m x_2 +10 k m +96 k x_1 -4 k x_2 \nonumber\\
& \qquad \qquad \qquad +2 k +6 m^2 -32 m x_1 +8 m x_2 -2 m -96 x_1 +8 x_2 -12 \Bigr) \ , \\ 
r_+ & = -\frac{1}{2 (k-2)^2} \Bigl( -5 j k-6 j m+4 j -2 k^2 x_2 -4 k m x_2 \nonumber \\
& \qquad \qquad \qquad \quad +5 k m +4 k x_2 +10 k+6 m^2+8 m x_2 +8 m-8 \Bigr) \ , \\ 
r_- & = -\frac{1}{2 (k-2)^2} \Bigl(8 j^2 k x_1 -16 j^2 x_1 -8 j k x_1 +11 j k+6 j m +16 j x_1 \nonumber\\
& \qquad \qquad \qquad \quad -10 j -4 k^3 x_1 -8 k^2 m x_1 -8 k m^2 x_1 -8 k m x_1 +11 k m \nonumber\\
& \qquad \qquad \qquad \quad +22 k+16 m^2 x_1 +6 m^2+48 m x_1 +2 m+32 x_1 -20 \Bigr) \ . 
\end{align}
\end{subequations}

\section{Generic spectral flow}
\label{app:generic-w}

In this Appendix we give some details about the calculation for the twisted sector null vectors from Section~\ref{sec:twisted}. We begin with evaluating the level $\frac{1}{w}$ null vector in terms of worldsheet variables 
\be 
\LL_{-\tfrac{1}{w}} \ket{\sigma_w} = \left( \mathcal{L}_{-\frac{1}{w}} - \LLm_{-\frac{1}{w}} \right) \ket{j,m}^{(w)} \ .
\ee 
The contribution due to the matter CFT vanishes, as 
\begin{align}
\LLm_{-\frac{1}{w}} \ket{j,m}^{(w)} &=   \oint_0 \mathrm{d}z\ \left( (\partial \gamma)^{-1} \gamma^{1-\frac{1}{w}} T^{\text{m}} \right)(z) \ket{j,m}^{(w)} \\
& \qquad + \dfrac{c^{\text{m}}}{12} \oint_0 \mathrm{d}z\ \gamma^{1 - \frac{1}{w}} \left( \frac{3}{2} \left(\partial^2 \gamma \right)^2 (\partial \gamma)^{-3} - \partial^3 \gamma (\partial \gamma)^{-2} \right) (z)\ket{j,m} \\
&=  \oint_0 \mathrm{d}z\ \dfrac{(z \tilde \gamma(z))^{1-\frac{1}{w}}}{\tilde \gamma(z) + z \partial \tilde \gamma(z)} T^{\text{m}} (z) \ket{j,m}^{(w)}\nonumber \\
&\qquad + \dfrac{c^{\text{m}}}{8} \oint_0 \mathrm{d}z\ \dfrac{\left(z \tilde \gamma(z) \right)^{1-\frac{1}{w}} \left( 2 \partial \tilde \gamma(z) + z \partial^2 \tilde \gamma(z) \right)^2}{\left( \tilde \gamma(z) + z \partial \tilde \gamma(z)  \right)^3} \ket{j,m}^{(w)} \nonumber\\
&\qquad - \dfrac{c^{\text{m}}}{12} \oint_0 \mathrm{d}z\ \dfrac{\left(z \tilde \gamma(z) \right)^{1-\frac{1}{w}} \left(3 \partial^2 \tilde \gamma(z) + z \partial^3 \tilde \gamma(z) \right)}{\left( \tilde \gamma(z) + z \partial \tilde \gamma(z)\right)^2}  \ket{j,m}^{(w)}=0\ ,
\end{align}
where the last equality follows from noticing that $\tilde \gamma(z)$ only produces regular terms when acting on $\ket{j,m}^{(w)}$ (see eq.~\eqref{eq:tilde-gamma-V-OPE}). 
On the other hand, we have from the definition of the DDF operators that 
\be  
\mathcal{L}_{-\frac{1}{w}} \ket{j,m}^{(w)} = \dfrac{1}{w} \oint_0 \mathrm{d}z\  \left( (w-1) \gamma^{-\frac{1}{w}} J^3 + \gamma^{1-\frac{1}{w}} J^+ \right)(z) \ket{j,m}^{(w)} \ . 
\ee
We first use Wick's theorem to disentangle the normal ordered product, and then express the $\gamma$-modes in terms of $\tilde{\gamma}$-modes. Then using eq.~\eqref{eq:tilde-gamman-V-OPE} (which we now apply for fractional powers of $\tilde{\gamma}$) we find
\be 
\oint_0 \mathrm{d}z\ \left(\gamma^{-\frac{1}{w}} J^3 \right)(z) \ket{j,m}^{(w)} = \tilde J^3_{-1} \Ket{j, m+\tfrac{1}{w}}^{(w)} - \left(\dfrac{m}{w} + \dfrac{k}{2} \right) \tilde \gamma_{-1}\Ket{j, m+\tfrac{1}{w}+1}^{(w)} \ . 
\ee
Similarly, by using eq.~\eqref{eq:gamma^n_0}, we find   
\begin{align}
\oint_0 \mathrm{d}z\ \left(\gamma^{-\frac{1}{w}+1} J^+ \right)(z) \ket{j,m}^{(w)} &= \dfrac{(w-1)(m+j-1)}{w} \tilde \gamma_{-1} \Ket{j, m + \tfrac{1}{w}+1}^{(w)} \nonumber\\
&\qquad +\tilde J^+_{-1} \Ket{j,m+\tfrac{1}{w}-1}^{(w)} \ . 
\end{align}
Collecting all the terms and using eq.~\eqref{eq:gamma-1} we thus obtain
\begin{align}
\LL_{-\frac{1}{w}}\ket{\sigma_w} = \mathcal{L}_{-\frac{1}{w}} \ket{j,m}^{(w)} & = \frac{(w-1) (-2 j+k w+2)}{2 w^2 (2 j+k-2)} \tilde J^-_{-1} \Ket{j, m+\tfrac{1}{w}+1}^{(w)} \nonumber\\
& \qquad + \frac{2 \left(w^2-1\right) (j-1)}{w^2 (2 j+k-2)} \tilde J^3_{-1} \Ket{j, m+\tfrac{1}{w}}^{(w)} \nonumber\\
& \qquad + \frac{(w+1) (2 j+k w-2)}{2 w^2 (2 j+k-2)} \tilde J^+_{-1} \Ket{j, m+\tfrac{1}{w}-1}^{(w)} \ . \label{eq:-1/wstate}
\end{align}


\begin{thebibliography}{99}

\bibitem{Maldacena:1997re}
J.M.~Maldacena,
``The Large N limit of superconformal field theories and supergravity,''
Int.\ J.\ Theor.\ Phys.\  {\bf 38} (1999) 1113
 [Adv.\ Theor.\ Math.\ Phys.\  {\bf 2} (1998) 231]
{\tt [hep-th/9711200]}.

\bibitem{Gopakumar:1998ki}
R.~Gopakumar and C.~Vafa,
``On the gauge theory / geometry correspondence,''
Adv.\ Theor.\ Math.\ Phys.\  {\bf 3} (1999) 1415
[AMS/IP Stud.\ Adv.\ Math.\  {\bf 23} (2001) 45]
  {\tt [hep-th/9811131]}.
  
\bibitem{Klebanov:2002ja}
I.R.~Klebanov and A.M.~Polyakov,
``AdS dual of the critical $\text{O}(N)$ vector model,''
Phys.\ Lett.\ B {\bf 550} (2002) 213
{\tt  [hep-th/0210114]}.
  
\bibitem{Maldacena:2000hw}
J.M.~Maldacena and H.~Ooguri,
``Strings in $\mathrm{AdS}_3$ and $\mathrm{SL}(2,\mathds{R})$ WZW model 1.: The Spectrum,''
J.\ Math.\ Phys.\  {\bf 42} (2001) 2929
{\tt [hep-th/0001053]}.

\bibitem{DiFrancesco:1997nk}
P.~Di Francesco, P.~Mathieu and D.~Senechal,
``Conformal Field Theory,''
Springer (1997).

\bibitem{David:2002wn}
  J.R.~David, G.~Mandal and S.R.~Wadia,
``Microscopic formulation of black holes in string theory,''
Phys.\ Rept.\  {\bf 369} (2002) 549
{\tt [hep-th/0203048]}.
        
\bibitem{Dijkgraaf:1998gf}
R.~Dijkgraaf,
``Instanton strings and hyperKahler geometry,''
Nucl.\ Phys.\ B {\bf 543} (1999) 545
{\tt [hep-th/9810210]}. 
  
\bibitem{Larsen:1999uk}
F.~Larsen and E.J.~Martinec,
``$\mathrm{U}(1)$ charges and moduli in the D1-D5 system,''
JHEP {\bf 9906} (1999) 019
{\tt [hep-th/9905064]}.   

\bibitem{Seiberg:1999xz}
N.~Seiberg and E.~Witten,
``The D1 / D5 system and singular CFT,''
JHEP {\bf 9904} (1999) 017
{\tt [hep-th/9903224]}.
  
\bibitem{Argurio:2000tb}
R.~Argurio, A.~Giveon and A.~Shomer,
``Superstrings on $\mathrm{AdS}_3$ and symmetric products,''
JHEP {\bf 0012} (2000) 003
{\tt [hep-th/0009242]}.  
  
\bibitem{deBoer:1998us}
J.~de Boer,
``Large $N$ elliptic genus and AdS / CFT correspondence,''
JHEP {\bf 9905} (1999) 017
{\tt [hep-th/9812240]}.    
  
\bibitem{Maldacena:1999bp}
J.M.~Maldacena, G.W.~Moore and A.~Strominger,
``Counting BPS black holes in toroidal Type II string theory,''
{\tt hep-th/9903163}.
  
\bibitem{Gaberdiel:2007vu}
M.R.~Gaberdiel and I.~Kirsch,
``Worldsheet correlators in $\mathrm{AdS}_3/\mathrm{CFT}_2$,''
JHEP {\bf 0704} (2007) 050
{\tt [hep-th/0703001]}.  
  
\bibitem{Dabholkar:2007ey}
A.~Dabholkar and A.~Pakman,
``Exact chiral ring of $\mathrm{AdS}_3/\mathrm{CFT}_2$,''
Adv.\ Theor.\ Math.\ Phys.\  {\bf 13} (2009)   409
{\tt [hep-th/0703022]}.  

\bibitem{Gaberdiel:2018rqv}
  M.R.~Gaberdiel and R.~Gopakumar,
  ``Tensionless string spectra on AdS$_{3}$,''
  JHEP {\bf 1805} (2018) 085
  {\tt [arXiv:1803.04423 [hep-th]]}.  
  
\bibitem{Eberhardt:2018ouy}
L.~Eberhardt, M.R.~Gaberdiel and R.~Gopakumar,
``The Worldsheet Dual of the Symmetric Product CFT,''
JHEP {\bf 1904} (2019) 103
{\tt [arXiv:1812.01007 [hep-th]]}.  

\bibitem{Gaberdiel:2014cha}
M.R.~Gaberdiel and R.~Gopakumar,
``Higher Spins \& Strings,''
JHEP {\bf 1411} (2014) 044
{\tt [arXiv:1406.6103 [hep-th]]}.
  
\bibitem{Gaberdiel:2017oqg}
M.R.~Gaberdiel, R.~Gopakumar and C.~Hull,
``Stringy AdS$_{3}$ from the worldsheet,''
JHEP {\bf 1707} (2017) 090
{\tt [arXiv:1704.08665 [hep-th]]}.
  
\bibitem{Ferreira:2017pgt}
K.~Ferreira, M.R.~Gaberdiel and J.I.~Jottar,
``Higher spins on AdS$_{3}$ from the worldsheet,''
JHEP {\bf 1707} (2017) 131
{\tt [arXiv:1704.08667 [hep-th]]}.

\bibitem{Giribet:2018ada}
G.~Giribet, C.~Hull, M.~Kleban, M.~Porrati and E.~Rabinovici,
``Superstrings on AdS$_{3}$ at $k=1$,''
JHEP {\bf 1808} (2018) 204
{\tt [arXiv:1803.04420 [hep-th]]}.
  
\bibitem{Eberhardt:2019qcl}
L.~Eberhardt and M.R.~Gaberdiel,
``String theory on $\text{AdS}_3$ and the symmetric orbifold of Liouville theory,''
{\tt arXiv:1903.00421 [hep-th]}.

\bibitem{Giveon:1998ns}
A.~Giveon, D.~Kutasov and N.~Seiberg,
``Comments on string theory on $\mathrm{AdS}_3$,''
Adv.\ Theor.\ Math.\ Phys.\  {\bf 2} (1998) 733
{\tt [hep-th/9806194]}.

\bibitem{deBoer:1998gyt}
J.~de Boer, H.~Ooguri, H.~Robins and J.~Tannenhauser,
``String theory on $\mathrm{AdS}_3$,''
JHEP {\bf 9812} (1998) 026
{\tt [hep-th/9812046]}.
  
\bibitem{Kutasov:1999xu}
D.~Kutasov and N.~Seiberg,
``More comments on string theory on $\mathrm{AdS}_3$,''
JHEP {\bf 9904} (1999) 008
{\tt [hep-th/9903219]}.

\bibitem{Belavin:1984vu}
A.A.~Belavin, A.M.~Polyakov and A.B.~Zamolodchikov,
``Infinite Conformal Symmetry in Two-Dimensional Quantum Field Theory,''
Nucl.\ Phys.\ B {\bf 241} (1984) 333.

\bibitem{Teschner:1995yf}
J.~Teschner,
``On the Liouville three point function,''
Phys.\ Lett.\ B {\bf 363} (1995) 65
{\tt [hep-th/9507109]}.

\bibitem{Dorn:1992at}
H.~Dorn and H.J.~Otto,
``On correlation functions for noncritical strings with c $\leq$ 1 d $\geq$ 1,''
Phys.\ Lett.\ B {\bf 291} (1992) 39
{\tt [hep-th/9206053]}.
  
\bibitem{Dorn:1994xn}
H.~Dorn and H.J.~Otto,
``Two and three point functions in Liouville theory,''
Nucl.\ Phys.\ B {\bf 429} (1994) 375
{\tt [hep-th/9403141]}.
  
\bibitem{Zamolodchikov:1995aa}
A.B.~Zamolodchikov and Al.B.~Zamolodchikov,
``Structure constants and conformal bootstrap in Liouville field theory,''
Nucl.\ Phys.\ B {\bf 477} (1996) 577
{\tt [hep-th/9506136]}.

\bibitem{Maldacena:2001km}
J.M.~Maldacena and H.~Ooguri,
``Strings in $\mathrm{AdS}_3$ and the $\mathrm{SL}(2,\mathds{R})$ WZW model. Part 3. Correlation functions,''
Phys.\ Rev.\ D {\bf 65} (2002) 106006
{\tt [hep-th/0111180]}.  

\bibitem{Teschner:1997ft}
J.~Teschner,
``On structure constants and fusion rules in the $\mathrm{SL}(2,\mathds{C}) / \mathrm{SU}(2)$ WZNW model,''
Nucl.\ Phys.\ B {\bf 546} (1999) 390
{\tt [hep-th/9712256]}.
  
\bibitem{Teschner:1999ug}
J.~Teschner,
``Operator product expansion and factorization in the $\mathrm{H}^+_3$ WZNW model,''
Nucl.\ Phys.\ B {\bf 571} (2000) 555
{\tt [hep-th/9906215]}.  

\bibitem{Giribet:2007wp}
G.~Giribet, A.~Pakman and L.~Rastelli,
``Spectral Flow in $\mathrm{AdS}_3/\mathrm{CFT}_2$,''
JHEP {\bf 0806} (2008) 013
{\tt [arXiv:0712.3046 [hep-th]]}.  

\bibitem{Giribet:2001ft}
G.~Giribet and C.A.~Nunez,
``Correlators in $\mathrm{AdS}_3$ string theory,''
JHEP {\bf 0106} (2001) 010
{\tt [hep-th/0105200]}.  
  
\bibitem{Ribault:2005ms}
S.~Ribault,
``Knizhnik-Zamolodchikov equations and spectral flow in $\mathrm{AdS}_3$ string theory,''
JHEP {\bf 0509} (2005) 045
{\tt [hep-th/0507114]}.  

\bibitem{DelGiudice:1971yjh}
E.~Del Giudice, P.~Di Vecchia and S.~Fubini,
``General properties of the dual resonance model,''
Annals Phys.\  {\bf 70} (1972) 378.

\bibitem{LorenzMatthiasRajesh}
L.~Eberhardt, M.R.~Gaberdiel and R.~Gopakumar,
\textit{to appear}.

\bibitem{Green:1987sp}
M.B.~Green, J.H.~Schwarz and E.~Witten,
``Superstring Theory. Vol. 1: Introduction,''
CUP (1987). 

\bibitem{Thielemans:1991uw}
K.~Thielemans,
``A Mathematica package for computing operator product expansions,''
Int.\ J.\ Mod.\ Phys.\ C {\bf 2} (1991) 787.
 
\bibitem{Ribault:2014hia}
S.~Ribault,
``Conformal field theory on the plane,''
{\tt arXiv:1406.4290 [hep-th]}.   

\bibitem{Lunin:2000yv}
O.~Lunin and S.D.~Mathur,
``Correlation functions for $M^N / S_N$ orbifolds,''
Commun.\ Math.\ Phys.\  {\bf 219} (2001) 399
{\tt [hep-th/0006196]}.
  
\bibitem{Pakman:2009zz}
 A.~Pakman, L.~Rastelli and S.S.~Razamat,
``Diagrams for Symmetric Product Orbifolds,''
JHEP {\bf 0910} (2009) 034
{\tt [arXiv:0905.3448 [hep-th]]}.

\bibitem{Eberhardt:2019niq}
  L.~Eberhardt and M.R.~Gaberdiel,
 ``Strings on $\text{AdS}_3 \times \text{S}^3 \times \text{S}^3 \times \text{S}^1$,''
  JHEP {\bf 1906} (2019) 035
  {\tt [arXiv:1904.01585 [hep-th]]}.

\bibitem{Poghosian:1996dw}
R.H.~Poghossian,
``Structure constants in the $\mathcal{N}=1$ superLiouville field theory,''
Nucl.\ Phys.\ B {\bf 496} (1997) 451
{\tt [hep-th/9607120]}. 

\bibitem{Giribet:2000fy}
G.~Giribet and C.A.~Nunez,
``Aspects of the free field description of string theory on $\mathrm{AdS}_3$,''
JHEP {\bf 0006} (2000) 033
{\tt [hep-th/0006070]}.

\bibitem{Hosomichi:2000bm}
K.~Hosomichi, K.~Okuyama and Y.~Satoh,
``Free field approach to string theory on $\mathrm{AdS}_3$,''
Nucl.\ Phys.\ B {\bf 598} (2001) 451
{\tt [hep-th/0009107]}.

\bibitem{Malikov:1986aa}
F.G.~Malikov, B.L.~Feigin and D.B.~Fuks,
``Singular vectors in Verma modules over Kac-Moody algebras,''
Funct.\ Anal.\ Appl.\ {\bf 20} (1986) 103.
          
\end{thebibliography}
\end{document}